\documentclass[10pt, preprint2]{aastex}

\usepackage{natbib}

\newcommand{\etal}{{et.~al.}}

\shorttitle{A Diagnostic Tool for Dark GRBs}
\shortauthors{Rol~\etal}

\begin{document}

\title{How Special Are Dark Gamma-Ray Bursts: A Diagnostic Tool}
\author{Evert~Rol\altaffilmark{1,2,3,4}, Ralph~A.~M.~J.~Wijers\altaffilmark{3}, Chryssa~Kouveliotou\altaffilmark{5,4}, Lex~Kaper\altaffilmark{3}, Yuki~Kaneko\altaffilmark{6}}

\altaffiltext{1}{Department of Physics and Astronomy, University of Leicester, University Road, Leicester LE1 7RH, United Kingdom; er45@star.le.ac.uk}
\altaffiltext{2}{INAF -- Osservatorio Astronomico di Trieste, Via Tiepolo 11, 34131 Trieste, Italy}
\altaffiltext{3}{Astronomical Institute, University of Amsterdam, Kruislaan 403, 1098 SJ Amsterdam, The Netherlands; rwijers@science.uva.nl, lexk@science.uva.nl}
\altaffiltext{4}{Institute for Nuclear Theory, University of Washington, Seattle, Washington 98195-1550, USA}
\altaffiltext{5}{NASA/MSFC, SD-50, 320 Sparkman Drive, Huntsville, AL 35805, USA, chryssa.kouveliotou@nasa.gov}
\altaffiltext{6}{University of Alabama in Huntsville, NSSTC, SD-50, 320 Sparkman Drive, Huntsville, AL 35805, USA, yuki.kaneko@msfc.nasa.gov}

\begin{abstract}

We present here a comprehensive study of the optical/near-infrared (IR) upper limits for gamma-ray bursts that have an X-ray afterglow. 
We have extrapolated the X-ray afterglows to optical wavelengths based on the physics of the fireball blast wave model \citep[e.g.][]{rees1992:mnras258, meszaros1997:apj476}, and compared these results with optical upper limits for a large sample of bursts. We find a small set of only three bursts out of a sample of 20 for which the upper limits are not compatible with their X-ray afterglow properties within the context of any blast wave model. This sparse sample does not allow us to conclusively determine the cause of this optical/near-IR deficit. Extinction in the host galaxy is a likely cause, but high redshifts and different afterglow mechanisms might also explain the deficit in some cases. We note that the three bursts appear to have higher than average gamma-ray peak fluxes. In a magnitude versus time diagram the bursts are separated from the majority of bursts with a detected optical/near-IR afterglow. However, two GRBs with an optical afterglow (one of which is highly reddened) also fall in this region with dark bursts, making it likely that dark bursts are at the faint end of the set of optically detected bursts, and therefore the dark bursts likely form a continuum with the bursts with a detected optical afterglow. 
Our work provides a useful diagnostic tool for follow-up observations for potentially dark bursts; applied to the events detected with the Swift satellite, it will significantly increase our sample of truly dark bursts and shed light upon their nature.

\end{abstract}


\section{Introduction}

After the discovery of the first gamma-ray bursts (GRBs)
\citep{klebesadel1973:apj182}, it took nearly 30 years before emission
at other wavelengths (afterglow) was discovered
\citep{costa1997:nature387, vanparadijs1997:nature386,
frail1997:nature389} and the distance scale was determined
\citep{metzger1997:nature387}. We know today that the main reasons for
this delay were the relatively rapid decay of the afterglows and the
inaccuracy of the initial burst localizations: (optical) searches for
afterglows have to be performed within one or two days after the burst
has been triggered in gamma rays and before the afterglow flux level
has decreased below the telescope detection limit. These searches have
to be performed with large aperture telescopes (requiring arc-minute sized
error boxes) or with sufficiently long exposure times on smaller
telescopes. Both options were rarely feasible for a long time, as
either the delay between the occurrence of the GRB and the
availability of its position was too large, or the size of the error
box of the GRB was too large to perform efficient follow-up
observations. The rapidly available and accurate positions provided initially by
the Wide Field Cameras (WFCs) on board the Italian-Dutch BeppoSAX satellite, and
later with HETE-II, combined with X-ray afterglow counterparts localized
 with the BeppoSAX Narrow Field Instruments
(NFIs), Chandra and XMM-Newton, made rapid follow-up
observations at longer wavelengths possible, with a typical error circle
diameter of the same order as the field-of-view of a large optical
telescope.

It was noted very soon, however, that despite rapid follow-up and deep imaging, some bursts escaped the detection of an optical counterpart \citep[e.g.,][]{groot1998:apj493}, and have been traditionally designated as 'dark bursts'. Notably, an optical afterglow was found in only 35\% of the BeppoSAX bursts for which an X-ray (2--10 keV) afterglow was detected \citep[see e.g.][]{depasquale2003:apj592}. Interestingly, however, eight out of ten bursts recently detected with the soft X-ray camera (SXC) on board HETE-2, have a detected optical or near-IR afterglow, increasing the detection rate to 80\%. This discrepancy in their dark bursts rates reflects rather the current improvement of the community response in GRB alerts, as well as today's enhanced observational capabilities, than an intrinsic difference in the properties of the GRBs detected by the two satellites.

Several explanations have been brought forward to address the absence of detectable optical flux, apart from adverse observing conditions: dark bursts could be intrinsically faint at optical wavelengths (possibly reflecting a difference in their central engine from `normal' GRBs); they could be at high redshift, which would result in their optical/near-IR light being suppressed due to hydrogen (Lyman-$\alpha$) absorption; or they could be heavily extincted by gas and dust.
The latter would not be unexpected, since the afterglows are located close to the center of light of their host galaxy \citep{bloom2002:aj123}. Furthermore, it is known now that at least a sub-class of GRBs originate from the core collapse of a massive star \citep{hjorth2003:nature423,stanek2003:apj519}. As these massive stars are very short-lived, they are expected not to travel far away from their birth place, which is most likely a very dusty environment. There are cases, however, where the host galaxy shows a high column density of neutral hydrogen, as expected for massive-star progenitors, but where strong upper limits are set on the reddening (and thus the dust content) in the host galaxy; e.g. for GRB\,020124 \citep{hjorth2003:apj597} and for GRB\,030323 \citep{vreeswijk2004:aa419}. In such cases, it is possible that X-ray dust destruction might free the way for the afterglow radiation to be still detected \citep{fruchter2001:apj563}.

It is important to note here that the current designation of a burst as `dark' may reflect either that no afterglow has been found by optical/near-IR searches, or that the optical counterpart is extremely faint initially and it declines very rapidly, thus precluding detection when observed later than one day after the burst \citep[see e.g.][]{fynbo2001:aa369, hjorth2002:apj576, berger2002:apj581}.
In the present study, however, we define as `dark', bursts that fall below an upper limit in their afterglow flux, which we derive using the fireball model. 
In this physical definition of dark bursts, the input for our calculations is obtained from observations in other wavelengths (X-ray or radio) where the afterglow was detected.
Of course, if there is no detection at any wavelength, this definition is not applicable, and one has to revert to the original empirical definition. Thus, to define our sample of `physically dark bursts', we have limited ourselves to those with a detected X-ray afterglow, from which we can predict the optical/near-IR fluxes.

The typical power-law decline of the X-ray and optical afterglows as well as the afterglow broad-band spectra obtained from observations at radio, millimeter, infrared, optical and X-ray wavelengths, show that the standard fireball model proposed as an explanation for these phenomena \citep{meszaros1997:apj476, wijers1997:mnras288}, is to a large extent acceptable. This standard model has been modified to account for the possible collimation of the outflow of the gamma-ray burst and its afterglow \citep[e.g.][]{rhoads1997:apj487, rhoads1999:apj525}, and for the expansion of the afterglow into a medium with a density gradient (e.g. a stellar wind; \citealt{chevalier1999:apj520}).
In general the model can be constrained by the available observations of a GRB afterglow. On the other hand, if observations in a particular wavelength interval are missing, it is possible to use the afterglow model to predict the flux (as a function of time) in this wavelength regime, which is what we have done below.
The outline of this paper is as follows: in Sect.~\ref{section:darkbursts-method} we describe the method used to constrain the optical/near-IR magnitudes expected from the X-ray afterglows. The results are analyzed in Sect.~\ref{section:darkbursts-results} and discussed in the context of the current ideas to explain dark bursts. Finally, we show how our results can be used as a diagnostic tool for future observations and studies of dark bursts.

\section{The method \label{section:darkbursts-method}}

\subsection{X-ray measurements \label{section:darkbursts-xrays}}

We have collected the X-ray afterglow observations from various sources (refereed journals, proceedings, GCN Circulars) and obtained a total of 20 X-ray afterglows with no optically detected burst, as well as 1 X-ray afterglow for an optically faint burst (GRB\,020322).
The available X-ray fluxes were first converted to the frequency corresponding to the observed energy band logarithmic average, instead of the usually quoted flux in the 2--10 or 0.2--10 keV range. This was done using e.g. $F_{2-10} = \int_{2}^{10} F_{\mathrm{av}} \times (\frac{E}{E_{\mathrm{av}}})^{\beta} dE$, where $\beta$ is the energy spectral index, $E_{\mathrm{av}}$ the logarithmic average and $F_{\mathrm{av}}$ the corresponding flux (e.g., for the 2--10 keV interval, $E_{\mathrm{av}} = 5$ keV). Since the quoted fluxes are generally an average of the flux in the observed time interval, we interpolated the flux to a logarithmic average $t_{\mathrm{av}}$ using the temporal decay index $\delta$ in a similar way. The choice of this spectral and temporal average ensures that our fluxes are not heavily dependent on the power-law indices $\beta$ and $\delta$.

In cases where no spectral index was reported, we have assumed a Crab-like spectrum. The error induced by this assumption is generally small, due to the choice of the logarithmic average as our pivot energy: for example, for a $F_{\mathrm{2-10}}$ flux of $5 \cdot 10^{-9} \mbox{erg cm}^{-2}\mbox{s}^{-1}$  and $\beta$ between 0.6 to 1.4, the $F_5$ flux varies only between $5.8 \cdot 10^{-10}$ and $6.4 \cdot 10^{-10} \mbox{erg cm}^{-2}\mathrm{s}^{-1} \mathrm{keV}^{-1}$.
We have assumed a 20\% error in the flux and the temporal or spectral indices of the X-rays if no error was listed. Table \ref{table:darkbursts-xrays} lists the X-ray afterglows which have been used in our analysis.
\placetable{table:darkbursts-xrays}

\subsection{Optical data \label{section:darkbursts-optical}}

Optical and near-IR data were largely obtained from the GCN Circulars, with additions from refereed journals and conference proceedings. The $R$-band upper limit for GRB\,000214 was obtained during our observing programme on the 2.2m ESO telescope at the La Silla observatory. The data were reduced in a standard fashion and calibrated using a set of Landolt standard stars.

All optical and near-IR magnitudes have been corrected for Galactic extinction using the COBE-DIRBE maps \citep{schlegel1998:apj500}. Corrected magnitudes have been converted to fluxes using the Vega flux values provided by \citet{fukugita1995:pasp107} for optical magnitudes and those by \citet{beckwith1976:apj208} for infrared observations. 

Where the quoted upper limit was the Digitized Sky Survey (e.g. ``we find no new sources down to the limit of the DSS''), we have assumed a limiting magnitude of 20. Unfiltered observations were left out of the analysis, unless the authors have given a corresponding filter and magnitude. Table \ref{table:darkbursts-upperlimits} lists the upper limits we have used for comparison with the X-ray afterglows.
This list has been compiled by using the strongest upper limits available for a specific burst. Also, for the more recent bursts, currently only reports in GCN and IAU Circulars exist, which should be considered temporary (this of course may be true for several of the older bursts as well).
\placetable{table:darkbursts-upperlimits}

\subsection{Predicting the optical flux by extrapolating the X-ray afterglows \label{section:darkbursts-extrapolation}}

We extrapolated the X-ray afterglow to the epoch and wavelength of the available optical and near-IR upper limits, as follows: we first obtained a range of power-law indices $p$ of the electron energy distribution from the 1$\sigma$ extremes of the X-ray power law temporal and spectral indices. To do this, we assumed eight cases in the standard afterglow model, which were derived from a combination of the following: the cooling frequency $\nu_{\mathrm{c}}$ was above or below the X-rays, the epoch of the X-ray observation was before or after the jet-break in the light curve, and the fireball was expanding into a constant or a stellar wind density medium. For convenience, we have listed these combinations and the dependence of $\delta$ and $\beta$ on $p$ in Table \ref{table:darkbursts-powerlaw-indices}. Some of these cases were ruled out by the fact that the two electron, inferred from the X-ray $\delta$ and $\beta$ values respectively, are incompatible with each other. We did not, however, limit the allowed range of $p$ to $p > 2$ (for $p \le 2$, one needs to put a high-energy cut-off on the electron distribution; see e.g. \citealt{panaitescu2001:apj560} or \citealt{bhattacharya2001:basi29}). From the obtained values for $p$, we extrapolated in each case the $+1\sigma$ and $-1\sigma$ X-ray flux to the epoch and frequency of the optical/near-IR upper limit. For this extrapolation, we chose the extreme possibilities, such as placing the cooling frequency at either the optical frequency or at the X-ray frequency, or the epoch of the jet-break right before or after the epoch of the X-ray detection. In this way, we obtain the maximum and minimum values at optical/near-IR frequencies of the extrapolated X-ray afterglow.
Sometimes such an extrapolation is ruled out by the standard afterglow model, and the result was ignored. This happens for example when $\nu_{\mathrm{c}}$ reaches $\nu_{\mathrm{x}}$ when extrapolating back in time, while $p$ was derived from the case where $\nu_{\mathrm{x}} < \nu_{\mathrm{c}}$.
\placetable{table:darkbursts-powerlaw-indices}

By choosing the lowest and highest extrapolated flux, we obtain the flux range in which one would expect an optical counterpart to be found. Any upper limit \emph{above} this range cannot be used to constrain the brightness of the optical afterglow.  Any upper limit \emph{below} this flux range will define this burst as physically `dark'. Upper limits found within the flux range will be called `grey', and are subject to further study. Of course, the optical afterglow of grey bursts could very well be below the extrapolation of the X-ray afterglow, but so far there is no observational evidence supporting this assumption.

In some cases, the electron index $p$ inferred from the available spectral and temporal index reached a value below 2. Since this index is generally found to be $> 2$ for detected optical afterglows (but see some of the results of \citet{panaitescu2001:apj560}, who find indices of $p \approx 1.5$ for some afterglows), we have also calculated the lowest possible extrapolation with $p$ fixed at 2. In most cases, however, the $p=2$ extrapolation is well below the lowest minimum extrapolation from the available indices.

In the case of GRB\,040223, a potential dark burst, the temporal and spectral indices of the X-ray afterglow \citep{tiengo2004:gcn2548} are incompatible with each other within the fireball model used here. We have listed the X-ray properties in Table \ref{table:darkbursts-xrays}, but have not extrapolated the X-ray flux to the optical waveband.

\section{Results \label{section:darkbursts-results}}

\begin{figure}

  \begin{minipage}{.45\columnwidth}
    \begin{center}
      (a)
      \includegraphics[width=\columnwidth]{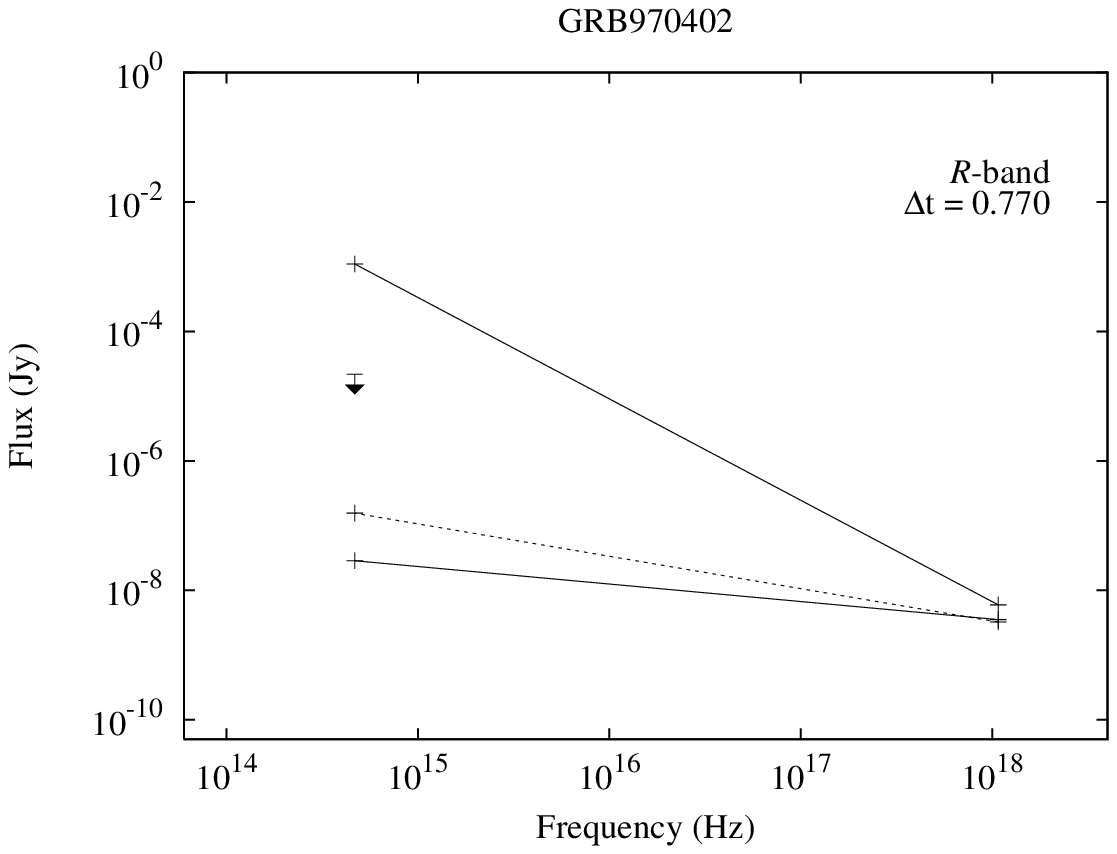}\\[5mm]
      (c)
      \includegraphics[width=\columnwidth]{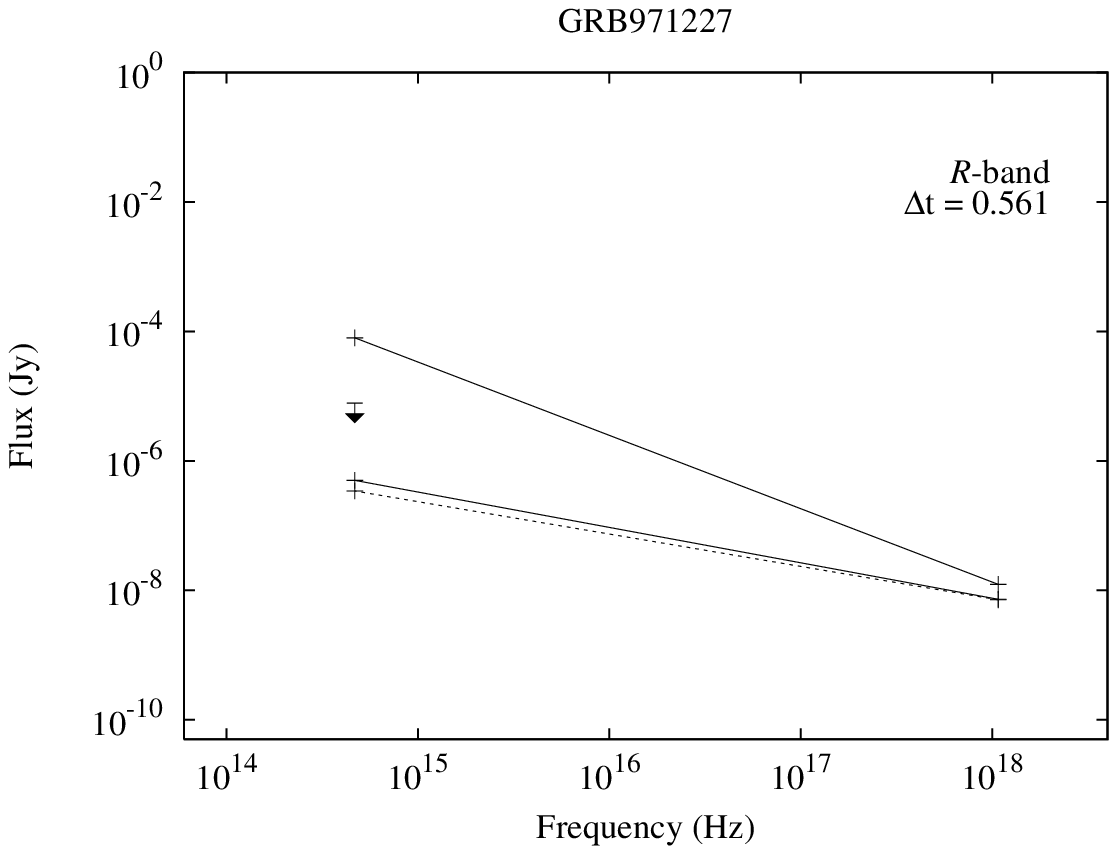}\\[5mm]
      (e)
      \includegraphics[width=\columnwidth]{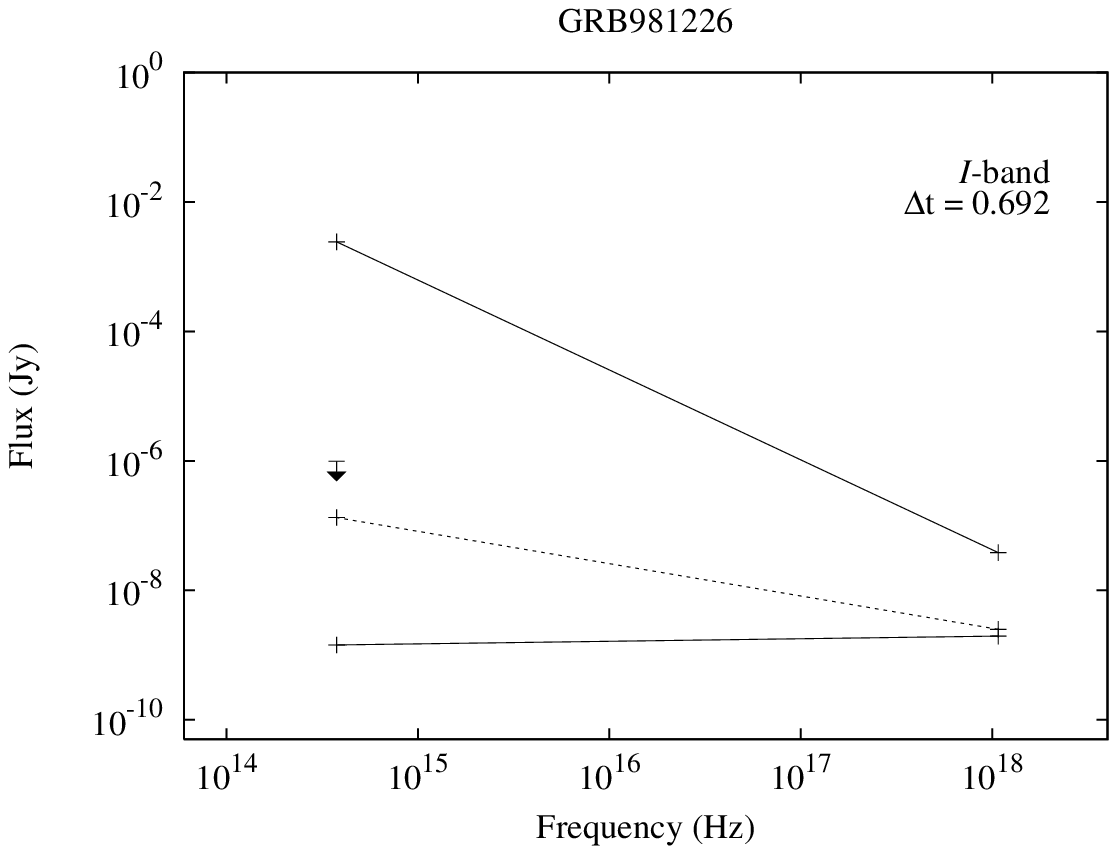}\\[5mm]
    \end{center}
  \end{minipage}
  \begin{minipage}{.45\columnwidth}
    \begin{center}
      (b)
      \includegraphics[width=\columnwidth]{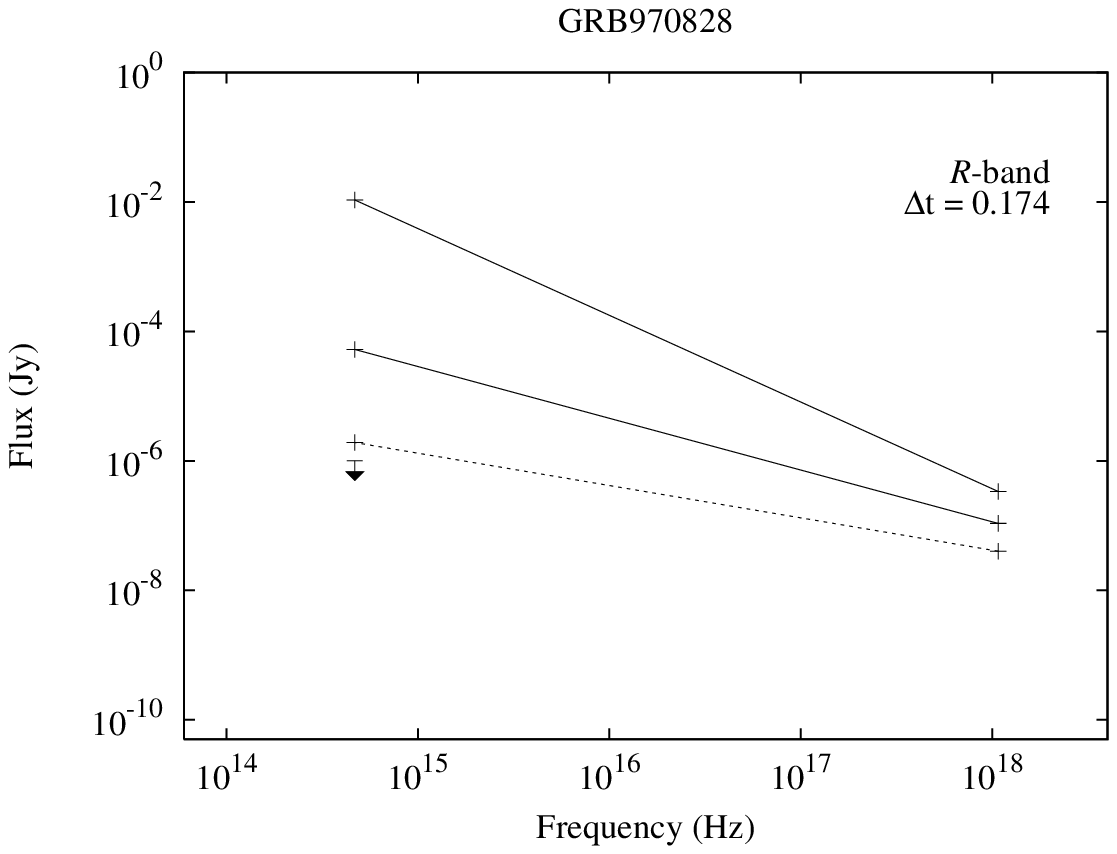}\\[5mm]
      (d)
      \includegraphics[width=\columnwidth]{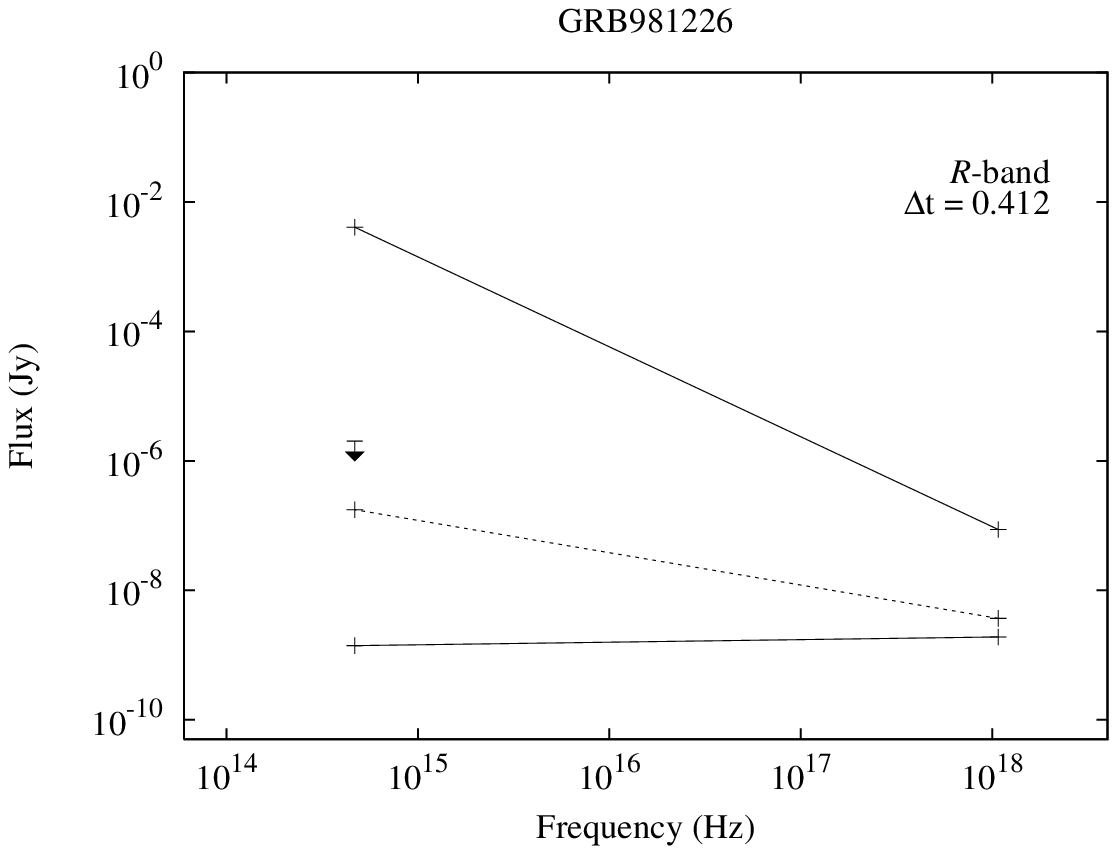}\\[5mm]
      (f)
      \includegraphics[width=\columnwidth]{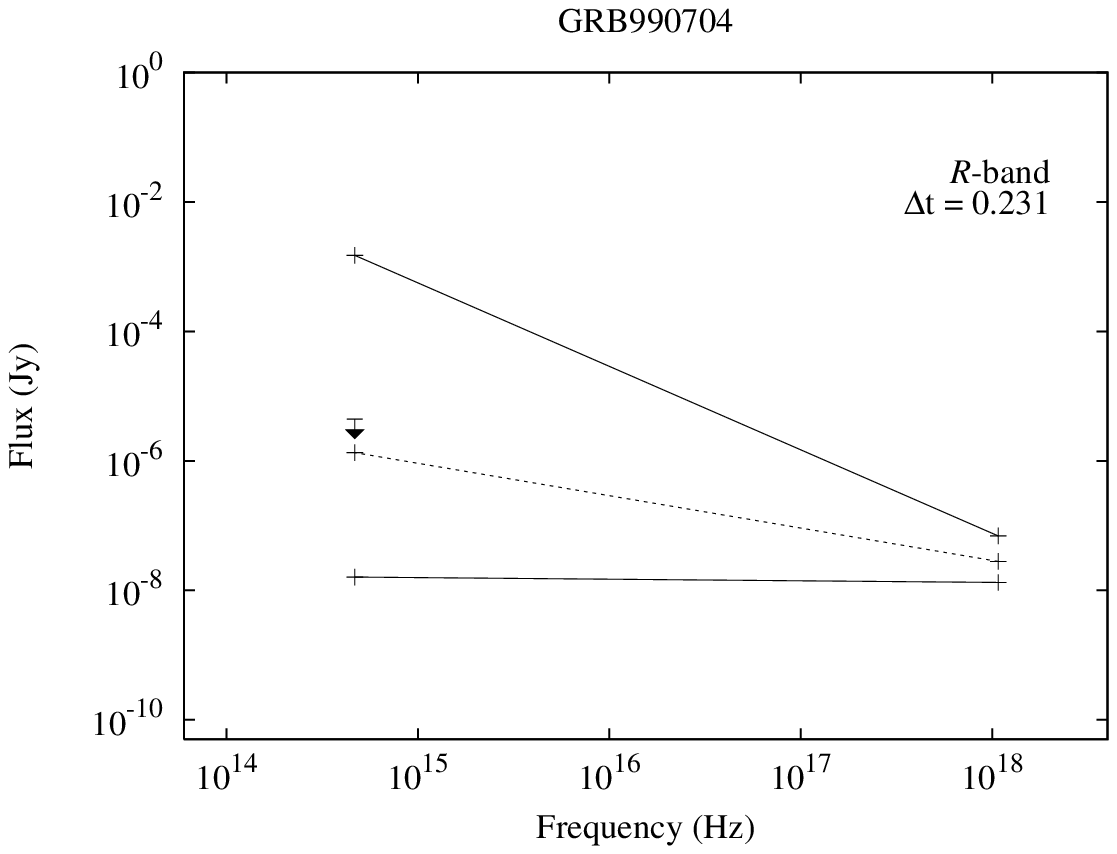}\\[5mm]
    \end{center}
  \end{minipage}

  \figurenum{1}
  \figcaption{\label{figure:darkbursts-xray-extrapolations} The ranges obtained for the optical counterpart fluxes for all GRBs in Table \ref{table:darkbursts-xrays}, by extrapolation of their X-ray measurements to the epoch and frequency of their optical upper limits; these limits are indicated by a downward-pointing arrow. 
The solid lines show the two extreme extrapolations, within which an optical counterpart should be found: the bottom line was obtained by extrapolating the $-1\sigma$ X-ray flux according to the lowest possible temporal and spectral slopes, and the top line was obtained by extrapolating the $+1\sigma$ X-ray flux using the highest possible temporal and spectral slopes. 
The dashed lines follow from extrapolating $-1\sigma$ X-ray flux with the lowest temporal and spectral slope with $p=2$.
}
 
\end{figure}

In Figure \ref{figure:darkbursts-xray-extrapolations} we show the results of the X-ray extrapolations and the various available upper limits at optical and near-IR wavelengths. 
For each X-ray afterglow, we only show the strongest optical and/or near-IR upper limit, as well as the minimum and maximum temporal extrapolation of the $-1 \sigma$ and $+1 \sigma$ X-ray flux to the frequency of the upper limit (solid lines). The dashed line is the extrapolation in the case $p = 2$, obtained for the $-1 \sigma$ X-ray flux and the minimum temporal and spectral extrapolation.

\placefigure{figure:darkbursts-xray-extrapolations}

There are only 3 GRBs out of the sample of 20 that we have found to be physically dark in this classification scheme: GRB\,970828, 000210 and 001025A. Two of these bursts have already been noted to be dark by extrapolation of the available X-ray measurements (GRB\,970828, \citealt{groot1998:apj493}, and GRB\,000210, \citealt{piro2002:apj577}).

In Table \ref{table:darkbursts-extinction}, we have indicated the dark bursts, together with the minimum amount of extinction (expressed in magnitudes) needed in the observer frame to explain the non-detection of the source. 
\placetable{table:darkbursts-extinction}

We have also plotted $R$-band magnitudes for all available afterglows together with all $R$-band upper limits in a magnitude--versus--time diagram (time being the delay between the GRB trigger and the observation; Figure \ref{figure:darkbursts-magtime}), and indicated which upper limits belong to the dark bursts. Upper limits in other bands with an available X-ray spectral index were converted to $R$-band magnitude following the same method as used in the X-ray extrapolations (i.e., the resulting $R$-band upper limit is the extrapolation with the least constraining magnitude). We have drawn a power law with temporal index $\alpha = -1.5$ passing through 25th magnitude at 2 days, which separates the three dark bursts from practically all detected afterglows. This power law is not based on any particular physical model, but merely serves as a guide in the diagram. 

The validity of this division, particularly at early times, has to be investigated, but at the current moment the lack of early X-ray afterglows prevents the study of dark bursts at early times in two ways: 1) early X-ray afterglows will generally provide more accurate positions than the initial GRB position, which gives larger telescopes, which often have a smaller field-of-view, the possibility to concentrate on deep searches, and 2) early X-ray afterglows will prevent the increase in the error when extrapolating the X-ray afterglow in the time-domain to its optical/near-IR upper limit.
This lack will be amended with the early afterglows available from Swift.

\placefigure{figure:darkbursts-magtime}

Two bursts with an optical afterglow fall below this line, GRB\,020322 and GRB\,030115. For GRB\,020322, the afterglow was detected at $R = 23.26$ \citep{bloom2002:gcn1294} after the identification of the X-ray afterglow with XMM-Newton \citep{ehle2002:gcn1293}. We have extrapolated the X-ray afterglow for this burst in the same way to its expected optical detection (see Figure \ref{figure:darkbursts-xray-extrapolations}z), and we find it to be compatible within the $1\sigma$ range of the X-ray extrapolation.
GRB\,030115 was $R = 24.3$ one day after the burst and is very red, with an $R-K$ color of about 6 mag (Levan \etal, ApJ submitted). Therefore, it is likely that the burst suffers from large extinction. Unfortunately, no X-ray afterglow has been reported for this burst. 

The other upper limits below this line belong to (in time order) GRB\,030324 \citep{moran2003:gcn2037} (first two upper limits), GRB\,021113 \citep{levan2002:gcn1688}, GRB\,030324 \citep{moran2003:gcn2037}, GRB\,000615 \citep{stanek2000:gcn709}, GRB\,990704 \citep{rol1999:gcn374}, GRB\,000214 (this paper), GRB\,001109 \citep{vreeswijk2000:gcn886}, GRB\,981226 \citep{lindgren1999:gcn190}, GRB\,990806 \citep[][equivalent $R$-band magnitude]{greiner1999:gcn396}, GRB\,991105 \citep{palazzi1999:gcn449}, GRB\,981226 \citep[][equivalent $R$-band magnitude]{bloom1998:gcn182} and GRB\,000830 \citep{jensen2000:gcn788}.
Seven of these bursts (GRB\,981226, GRB\,990704, GRB\,990806, GRB\,991105, GRB\,000214, GRB\,001109 and GRB\,000615) are included in our sample of analyzed afterglows, and shown in Figure \ref{figure:darkbursts-xray-extrapolations}. For two of these (GRB\,000214 and GRB\,001109), the optical/near-IR limit is very close to the minimum extrapolation from X-rays, putting them at the border of physically defined grey/dark bursts. In fact, as this minimum extrapolation is an extreme possibility for the afterglow behavior, it would not be unreasonable to qualify GRB\,000214 and GRB\,001109 as dark bursts. However, to adhere to our definition of physically dark bursts, we choose not do so here.

\section{Discussion \label{section:darkbursts-discussion}}

Figure \ref{figure:darkbursts-magtime} shows that in an $R$-magnitude--versus--time diagram, the physicallly defined dark bursts are located in a distinct region at the faint end of optically detected bursts.
The detection of GRB\,020322 and GRB\,030115, whose magnitudes are comparable to the upper limits of the two dark bursts GRB\,000210 and GRB\,001025A, suggests that dark and detected bursts form a continuous group, rather than two clearly separated samples, and that the afterglows of most of the dark burst candidates might be found by deeper searches. Such afterglows may very well be highly reddened, as is the case for GRB\,030115, and would require searches to be performed in the near-IR rather than in the optical wavebands.
One would expect that some dark bursts occupy the region of detected bursts; unfortunately, the X-ray data do not allow us to constrain the upper limits in this sample enough to find dark bursts in this region. Specifically, a well defined spectral index will constrain the result of the X-ray extrapolation to optical/near-IR wavelengths. Such a constrained X-ray extrapolation, combined with an optical upper limit, can show if physically dark bursts also occur in the upper part of Figure \ref{figure:darkbursts-magtime}.
More significantly, our method can be used as a diagnostic tool for the overall determination of dark bursts, provided that early X-ray and optical observations are available, to compare the calculation of the extrapolated values to the observed limits.

Here, we speculate on three possible causes for the (physical) darkness: high-redshift bursts, different afterglow properties and extinction within the host galaxy. However, we wish to note that our sample of three bursts is too small to draw any definite conclusions.

High-redshift bursts will go undetected at optical wavelengths due to extinction by the Ly$\alpha$ forest, although this would require an X-ray afterglow which is intrinsically much harder, as its redshifted flux is still detected in the 2--10 keV X-ray range. Furthermore, two dark bursts have a redshift from the detection of their host galaxy (GRB\,970828, $z = 0.9578$ \citep{djorgovski2001:apj562}, and GRB\,000210, $z = 0.846$ \citep{piro2002:apj577}), which is well within the sample of detected afterglows and does not suggest a high-redshift origin for dark bursts. The redshifts of the two detected bursts with very dim afterglows, which fall within our region of dark bursts (GRB\,020322 and GRB\,030115), are not known. However, for GRB\,030115 the possible host is visible at $R = 24.5$ \citep{garnavich2003:gcn1848}, putting its redshift at $z < 4.5$.

Dark bursts could form a subset of GRB afterglows with different afterglow properties. \citet{depasquale2003:apj592} have found that the X-ray fluxes of dark bursts are different than the X-ray fluxes of optically detected bursts. Their sample of dark bursts, however, is not physically defined as in this current study, but uses the observational definition, which includes non-detections that might be attributed to observing conditions.

To see if there is anything remarkable in the prompt gamma-ray properties, we have compared the 40-700 keV gamma-ray fluences of the three phsyically dark bursts with a general set of bursts, obtained from the BeppoSAX GRBM data and the BATSE data (the BATSE 40-700 keV fluences were obtained from spectral fits to the BATSE archival data, while the GRBM data were obtained from literature). 
This comparison set was selected by choosing bursts for which narrow error boxes are available and have had optical follow-up campaigns, to make the comparison set similar to the set of GRB data of the physically dark bursts.
The set includes among others all the grey bursts.
 For GRB\,001025A, the Ulysses 25-100 keV fluence was converted to the 40-700 keV range, using a conversion factor obtained by converting a set of 28 BATSE burst templates with known spectral shapes \citep[cf.][]{bloom2001:aj121}. The result is shown at the top of Figure \ref{figure:darkbursts-darkpeakflux/fluences}. No obvious trend is visible, although we note that the dark bursts appear to have a higher than average 40-700 keV fluence. This is also noticeable in the bottom figure, where we have plotted the peak fluxes of the dark bursts together with those for a general sample of bursts: the three dark bursts are in the higher part of the sample. However, our set is too small to meaningfully apply any formal statistical test, and we note it here as a possible trend. 
\placefigure{figure:darkbursts-darkpeakflux/fluences}

Such a correlation between the peak flux and the faintness of the optical afterglow could happen in the case of an external scenario for the GRB and its afterglow: if the prompt gamma-ray emission is the result of the interaction with an external medium instead of internal shocks, a higher density of the medium would then result in a higher gamma-ray peak flux of the burst. A higher density would, however, also result in a higher absorption depth (provided the path length is long enough) and the optical/UV emission would go undetected.

Extinction as a likely cause for dark bursts was already suggested by \citet{groot1998:apj493} for GRB\,970828. More recently, \citet{klose2003:apj592} explored the available $K$-band upper limits of GRBs and also found evidence for extinction to explain dark bursts. The two dark bursts with an identified host galaxy have already been discussed before. GRB\,970828 has one of the deepest early upper limits available, as well as a host galaxy and a redshift. \citet{djorgovski2001:apj562} show that a possible dust lane or giant molecular cloud could provide the amount of extinction needed to obscure the afterglow. For GRB\,000210, the X-ray spectrum shows significant $N_{\rm H}$ absorption; the amount of dust needed to obscure an optical afterglow at the redshift of the host galaxy is consistent with this column density. Both findings favor dust extinction in the host galaxy for these two dark bursts.

Extinction should also be noticeable in optically detected bursts, and the $R$-$K$ color of 6 mag for GRB\,030115 indeed suggests that this burst is heavily extincted. However, only very few bursts indeed show large extinction in the host galaxy. Even some notably faint afterglows, like GRB\,980613 \citep{hjorth2002:apj576} and GRB\,000630 \citep{fynbo2001:aa369}, have a low ($< 0.5$ magnitude) extinction value in the $R$-band (in addition to galactic extinction). It is therefore quite possible that we are simply not finding many heavily extincted bursts because these are still below the generally reported detection limits. However, the last 10 optical/near-IR follow-up campaigns of HETE-2 SXC error boxes resulted in 8 detected afterglows. This suggests that for the majority of the 60\%\ of the cases where no afterglow was found for a BeppoSAX GRB, this was due to adverse observing conditions. Our sample and the SXC sample suggest a fraction of physically dark bursts more like 10--20\%.
This may indicate that the few dark bursts we find here, are indicative of the size of the actual (physical) dark burst sample.

\section{Conclusions \label{section:darkbursts-conclusions}}

The extrapolation from the X-ray afterglow to optical/near-IR wavelengths shows that only very few bursts deserve the designation `dark burst'. The optical upper limits obtained for these dark bursts are within a region at the fainter end of the optically detected bursts, with two optical afterglows found within this region. One of these two optical afterglows is highly reddened, which suggests that dark bursts can indeed be detected, albeit at very faint magnitudes. Detection of such dark bursts provides the necessary insights into the cause of their darkness.

Our method of extrapolation provides a useful diagnostic tool for setting a lower limit on the expected magnitude of the optical counterpart. When an early X-ray light curve and spectrum are available, for example from the XRT on board Swift, one can set a `dark burst upper limit'. Any afterglow not detected above this upper limit is a potential candidate for a high redshift or an extinguished afterglow, or possibly even an afterglow produced by different physical mechanisms. Setting a search area using the currently available afterglow upper limits, as in Figure \ref{figure:darkbursts-magtime}, gives future afterglow searches a tool for picking out such potential dark bursts even without an X-ray afterglow.

It is, however, not possible to draw definite conclusions on the cause of darkness, due to the small set of dark bursts we obtained. We note that in the gamma-ray properties, dark bursts appear to have higher than average fluences and peak fluxes, but we cannot statistically quantify this result yet. We also find that at least one burst showing large reddening (GRB\,030115), is within the same region of magnitude--versus--time diagram (Figure \ref{figure:darkbursts-magtime}) as the dark bursts. 

Deep upper limits within one day, as well as early localizations and rapid follow-up observations will complete and enhance Figure \ref{figure:darkbursts-magtime}. The Swift mission can certainly help filling the gap for early times: it will provide early X-ray afterglow spectra and light curves, as well as  rapid and precise localizations, needed for early (robotic) searches and deep searches with larger telescopes.

\acknowledgements

ER acknowledges support from NWO grant nr. 614-51-003, and from PPARC grant nr PPa/G/D/2003/00018. Part of this analysis was performed while ER was visiting NSSTC in Huntsville under grant HSTG008189.05A. ER would also like to thank the hospitality of the Osservatorio Astronomico di Padova. 
The gamma-ray burst coordinate network (GCN) and its archive of corresponding circulars at \url{http://gcn.gsfc.nasa.gov/selected.html}, set-up and maintained by Scott Barthelmy, served as an invaluable basis for retrieving the enormous amount of upper limits found in GRB afterglow searches over the past six years. The web page at \url{http://www.mpe.mpg.de/$\sim$jcg/grbgen.html}, maintained by Jochen Greiner, served as an excellent cross-check for this compilation, and sometimes provided details not found elsewhere. This research has made extensive use of NASA's Astrophysics Data System Bibliographic Services. We like to thank Jens Hjorth for useful comments on the manuscript, and Ed van den Heuvel for useful discussions.
 The authors acknowledge benefits from collaboration within the Research Training Network "Gamma-Ray Bursts: An Enigma and a Tool", funded by the EU under contract HPRN-CT-2002-00294. 
We thank the referee for detailed comments on the manuscript.

\setlength{\bibsep}{0cm}
\bibliographystyle{apj}
\bibliography{references}

\onecolumn

\begin{figure}

  \begin{minipage}{.45\columnwidth}
    \begin{center}
      (g)
      \includegraphics[width=\columnwidth]{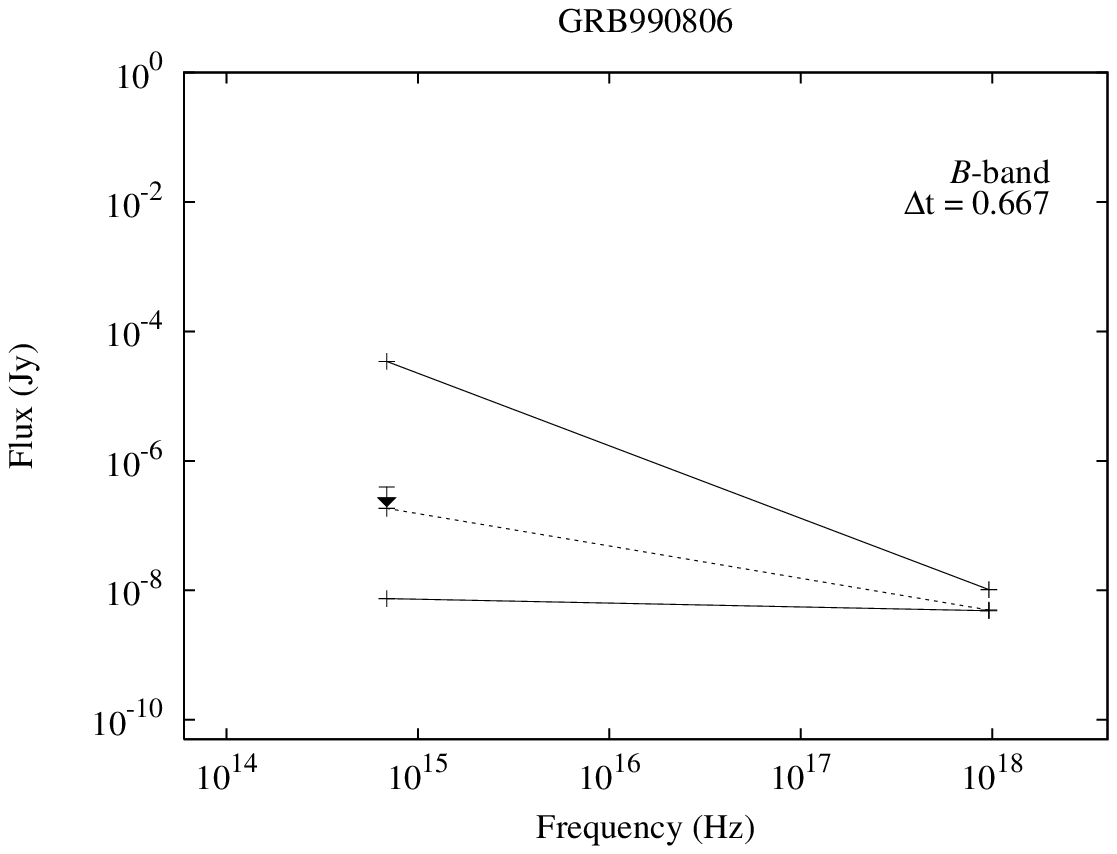}\\[5mm]
      (i)
      \includegraphics[width=\columnwidth]{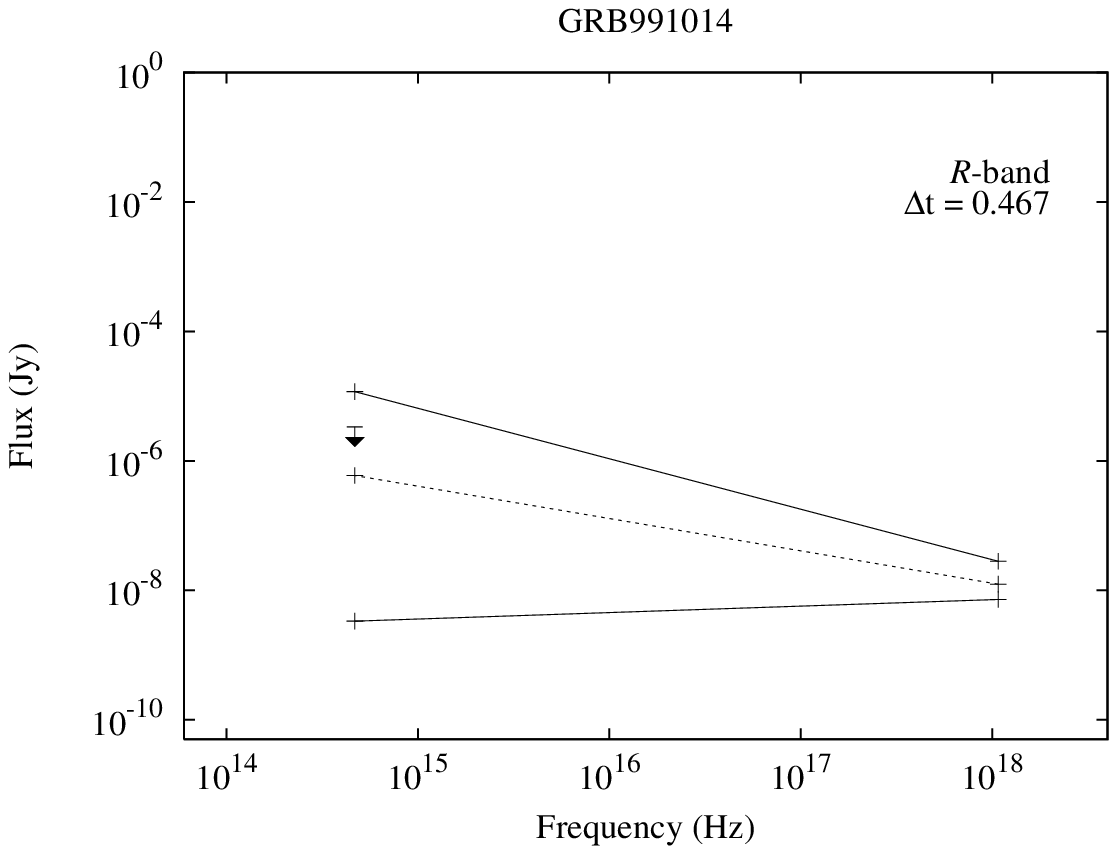}\\[5mm]
      (k)
      \includegraphics[width=\columnwidth]{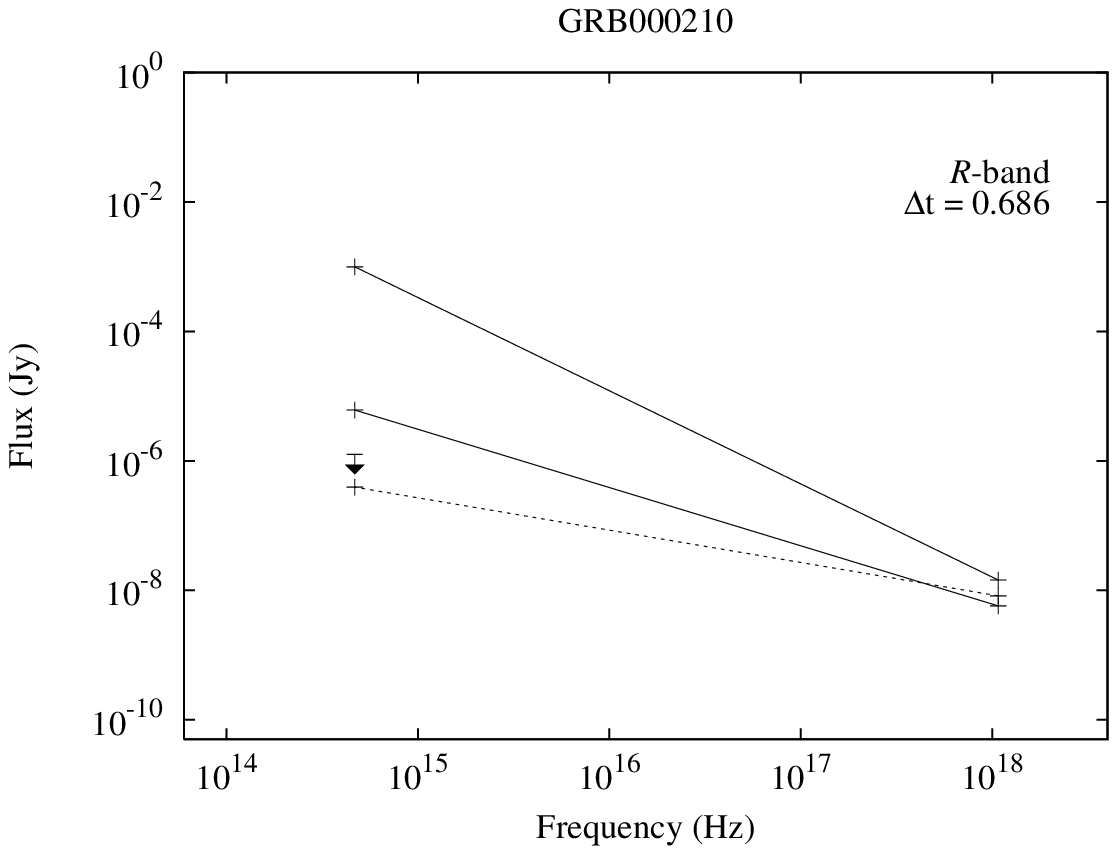}\\[5mm]
    \end{center}
  \end{minipage}
  \begin{minipage}{.45\columnwidth}
    \begin{center}
      (h)
      \includegraphics[width=\columnwidth]{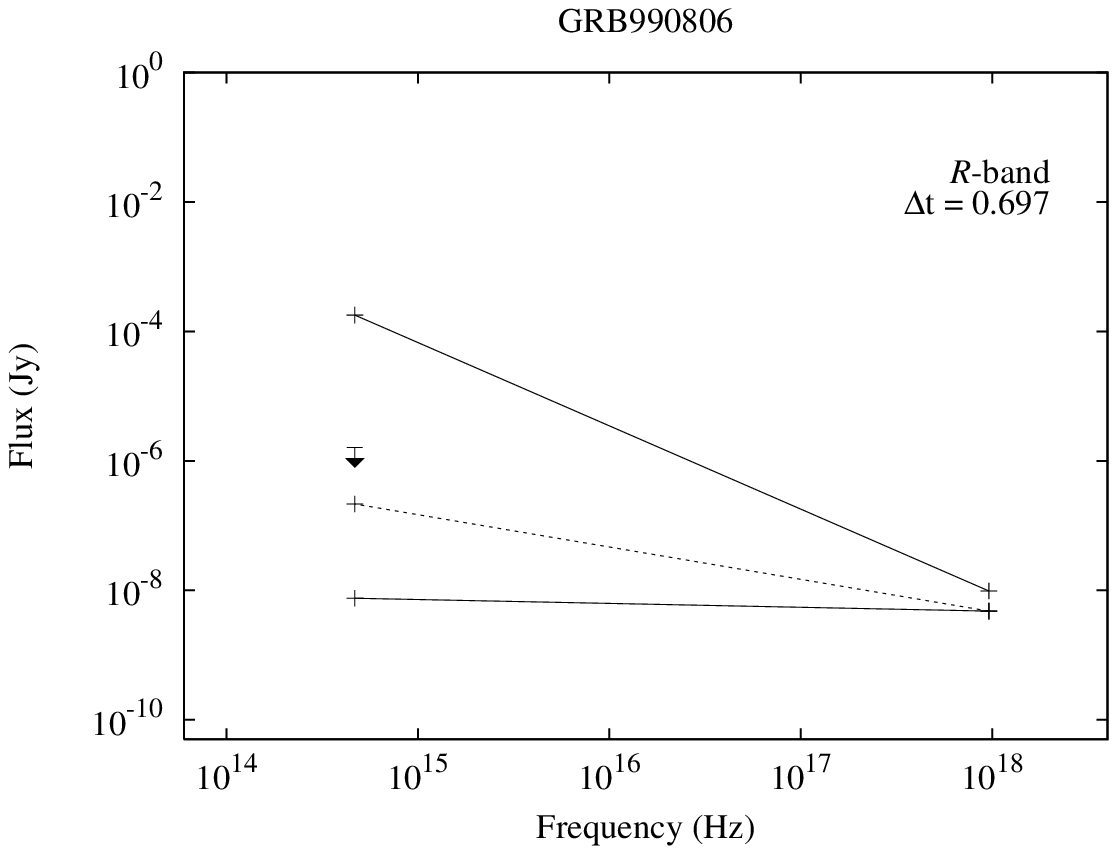}\\[5mm]
      (j)
      \includegraphics[width=\columnwidth]{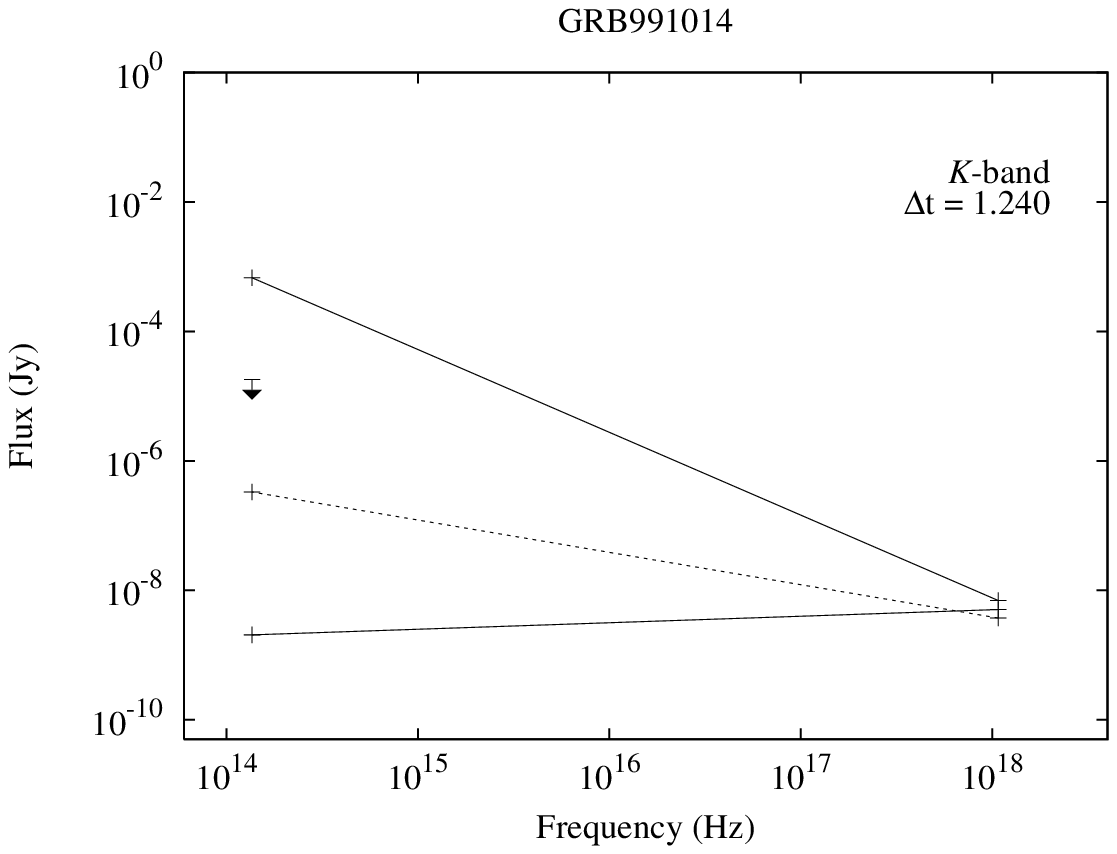}\\[5mm]
      (l)
      \includegraphics[width=\columnwidth]{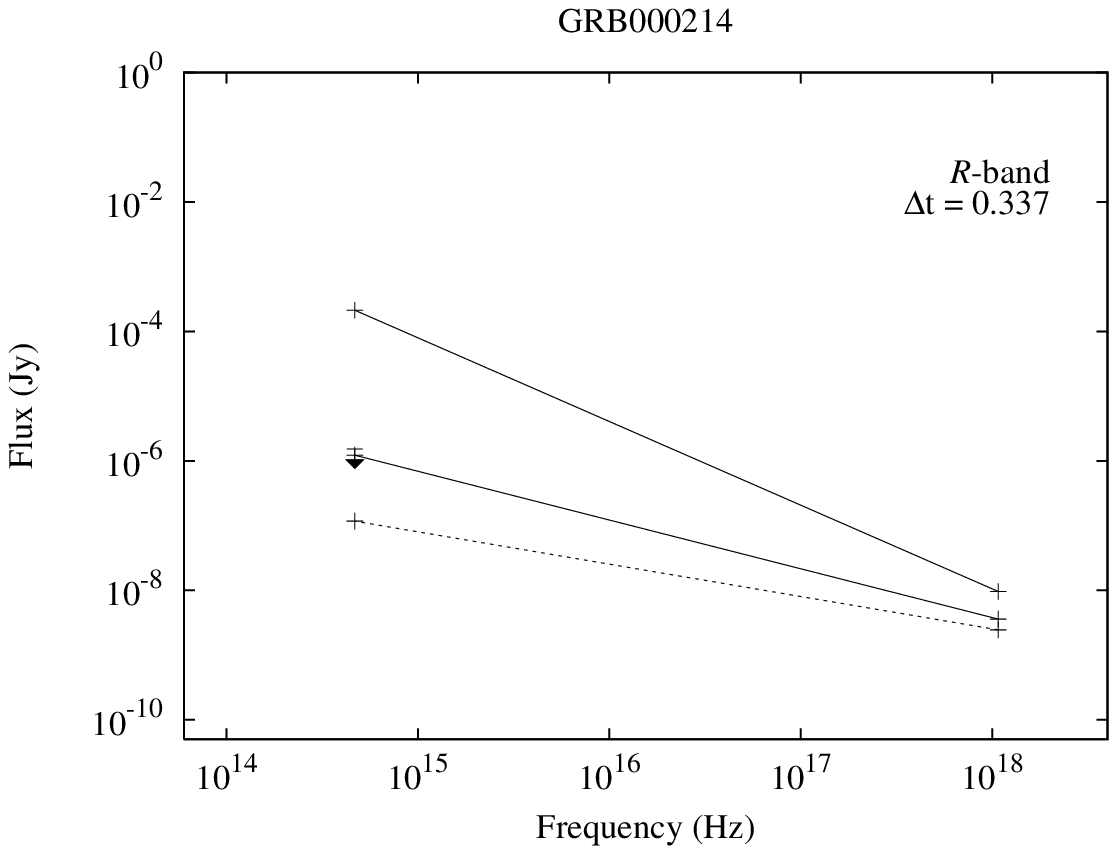}\\[5mm]
    \end{center}
  \end{minipage}

  \figurenum{1}
  \figcaption{(continued)}
 
\end{figure}

\begin{figure}

  \begin{minipage}{.45\columnwidth}
    \begin{center}
      (m)
      \includegraphics[width=\columnwidth]{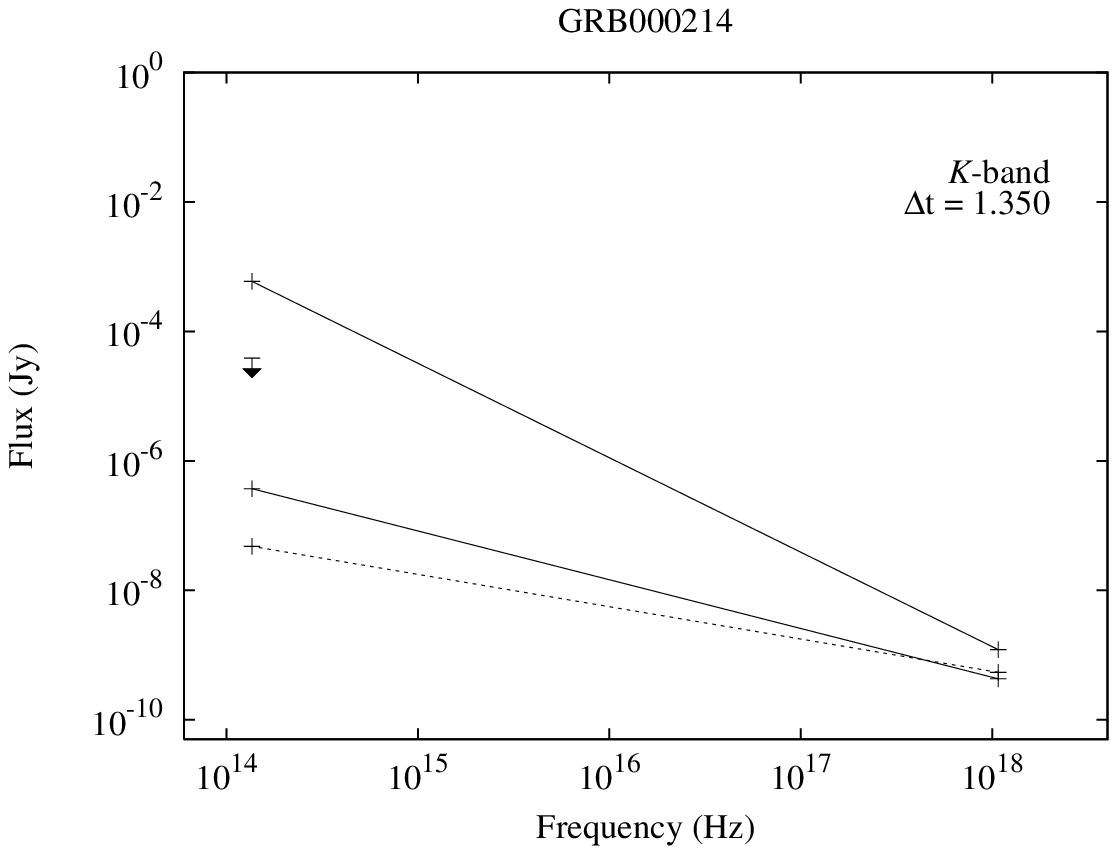}\\[5mm]
      (o)
      \includegraphics[width=\columnwidth]{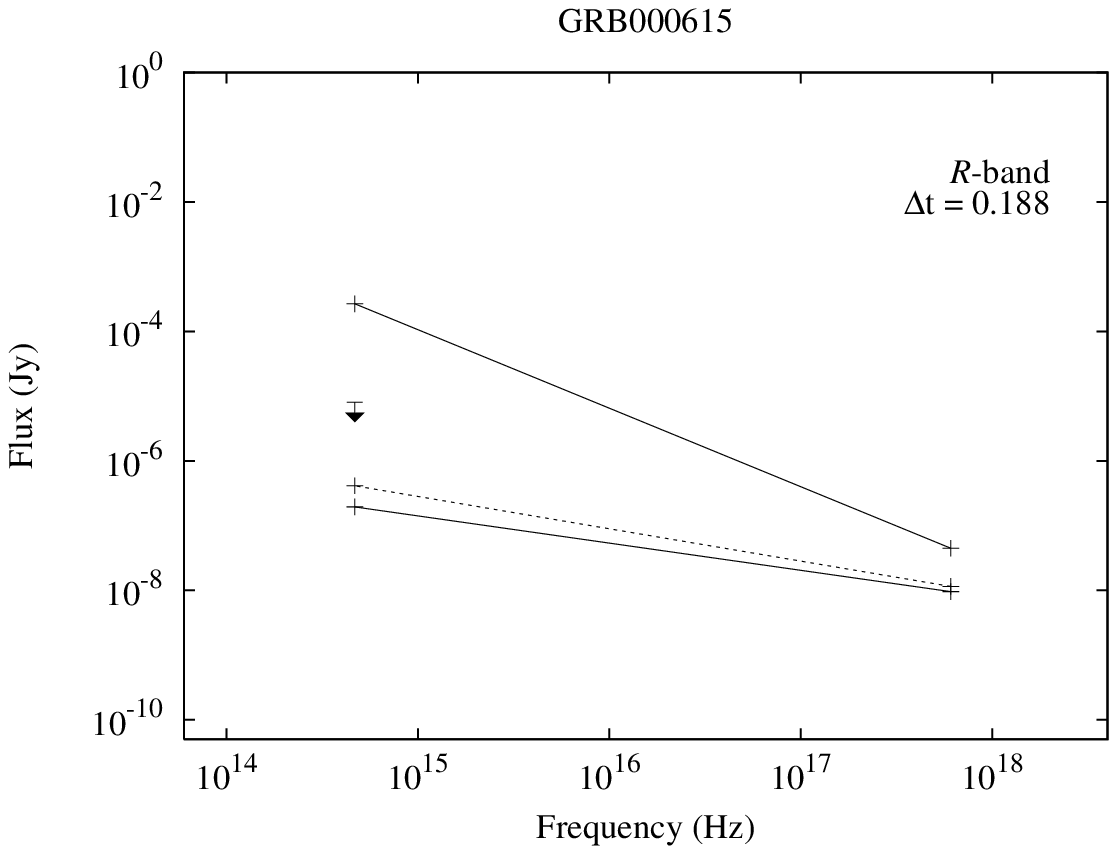}\\[5mm]
      (q)
      \includegraphics[width=\columnwidth]{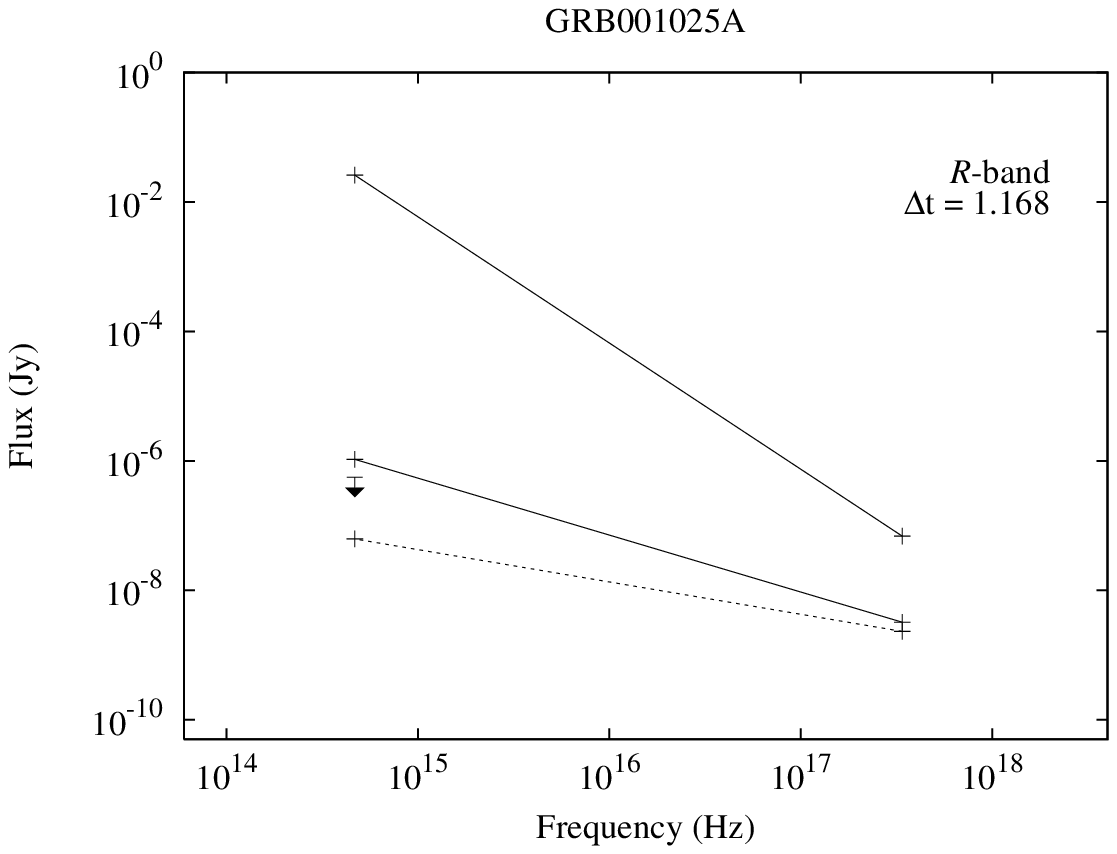}\\[5mm]
    \end{center}
  \end{minipage}
  \begin{minipage}{.45\columnwidth}
    \begin{center}
      (n)
      \includegraphics[width=\columnwidth]{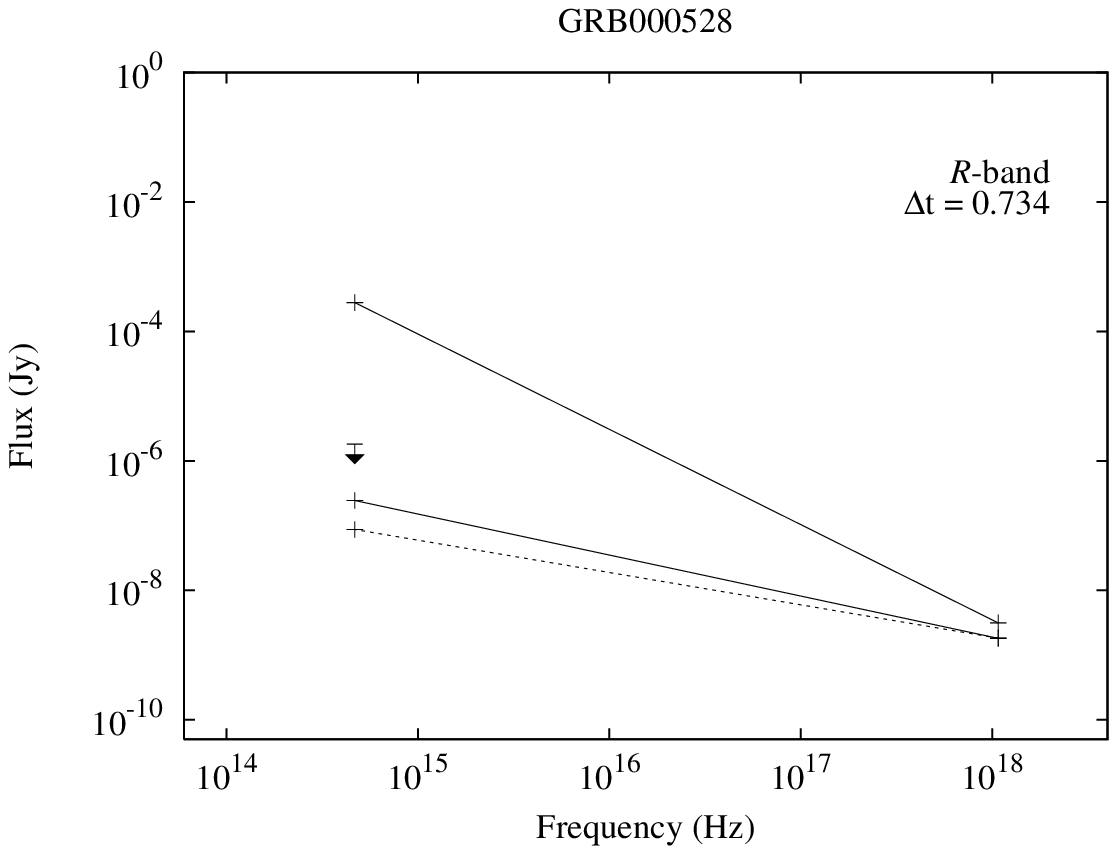}\\[5mm]
      (p)
      \includegraphics[width=\columnwidth]{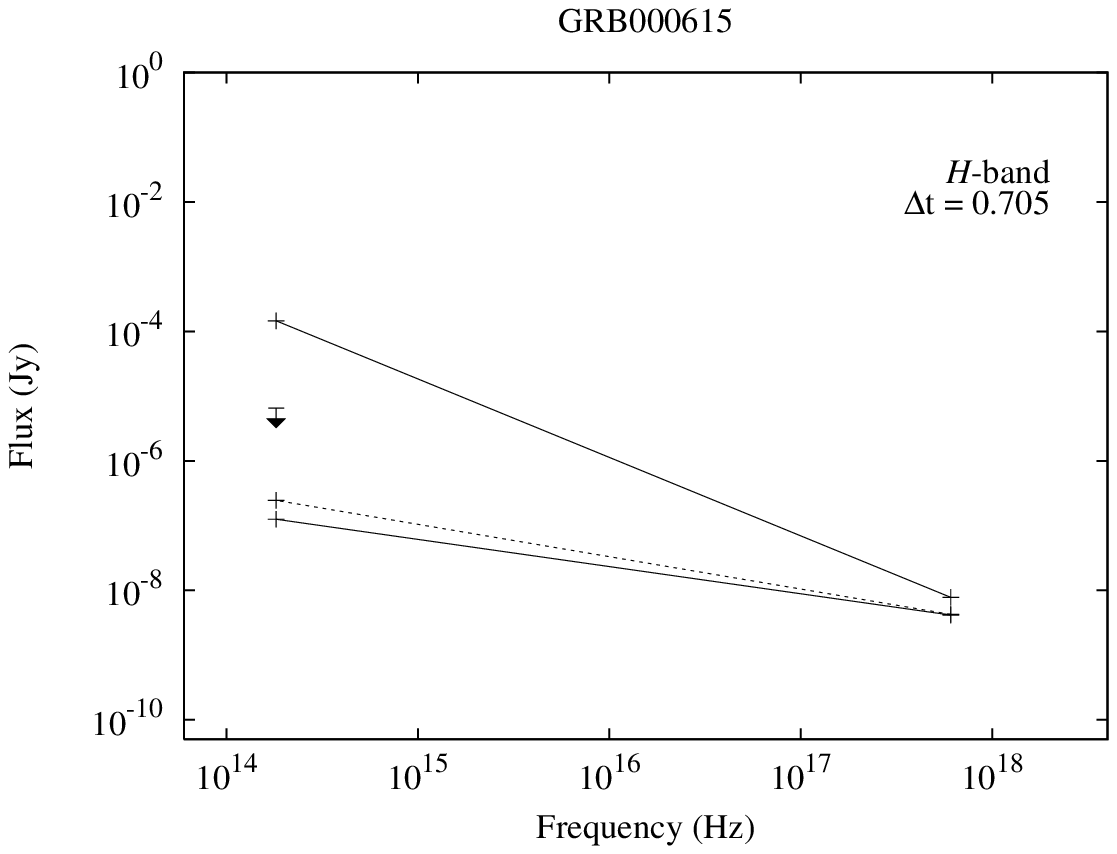}\\[5mm]
      (r)
      \includegraphics[width=\columnwidth]{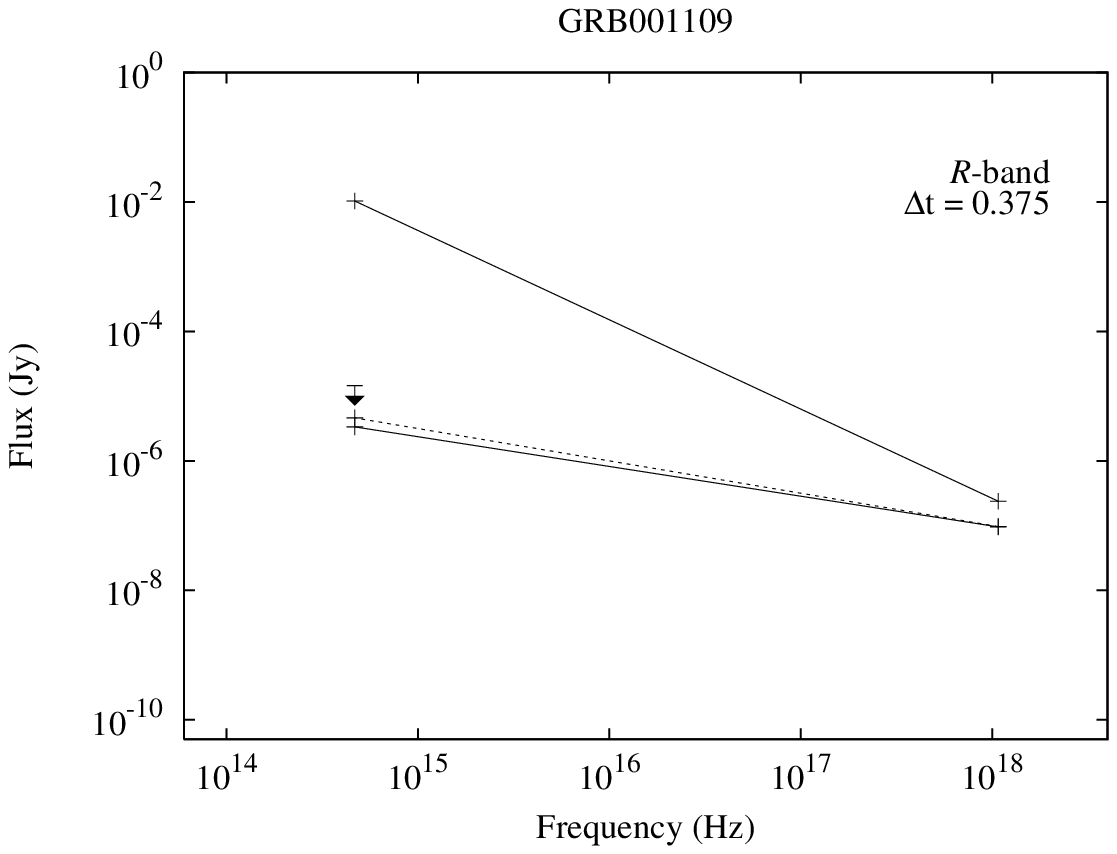}\\[5mm]
    \end{center}
  \end{minipage}

  \figurenum{1}
  \figcaption{(continued)}
 
\end{figure}

\begin{figure}

  \begin{minipage}{.45\columnwidth}
    \begin{center}
      (s)
      \includegraphics[width=\columnwidth]{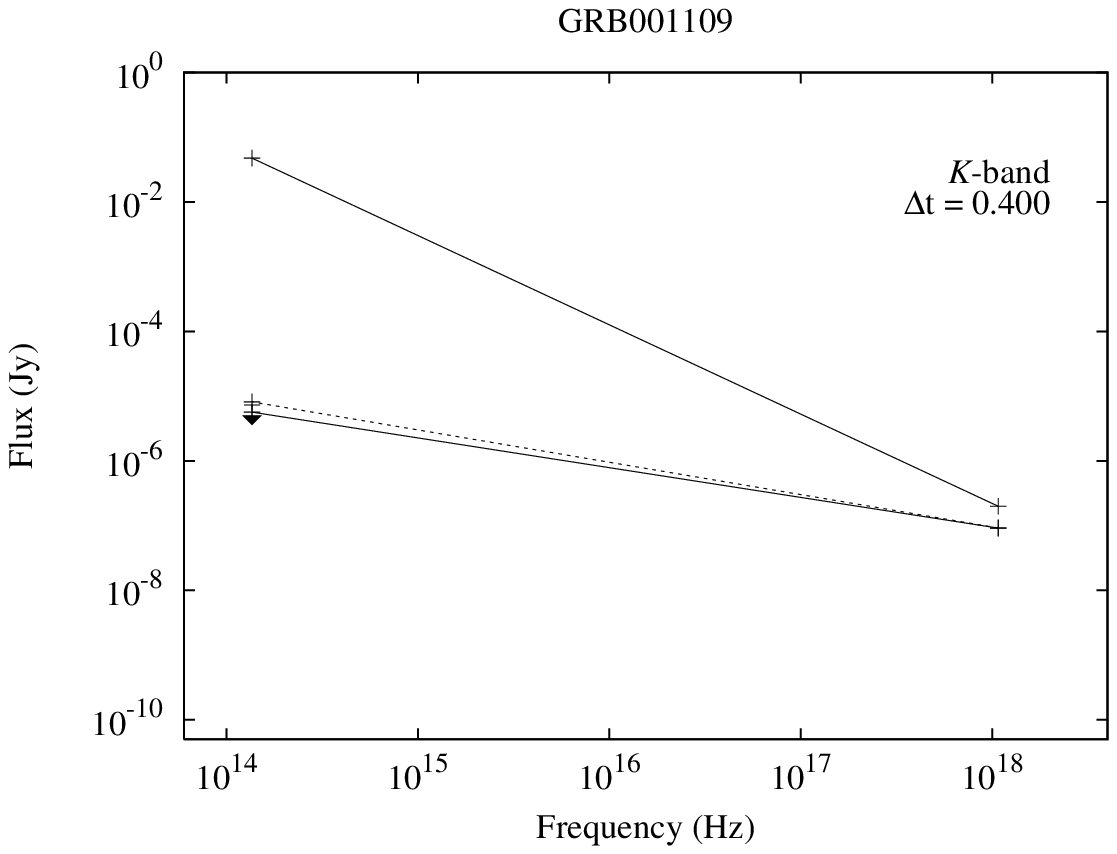}\\[5mm]
      (u)
      \includegraphics[width=\columnwidth]{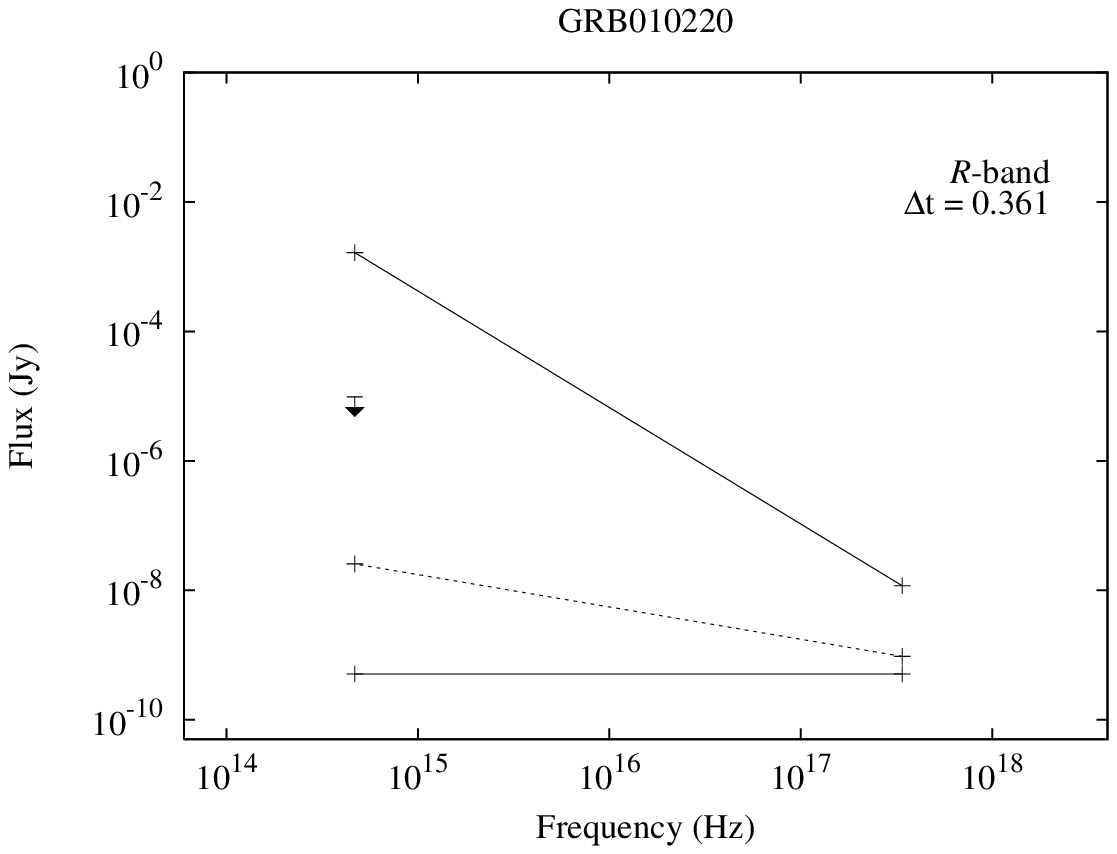}\\[5mm]
      (w)
      \includegraphics[width=\columnwidth]{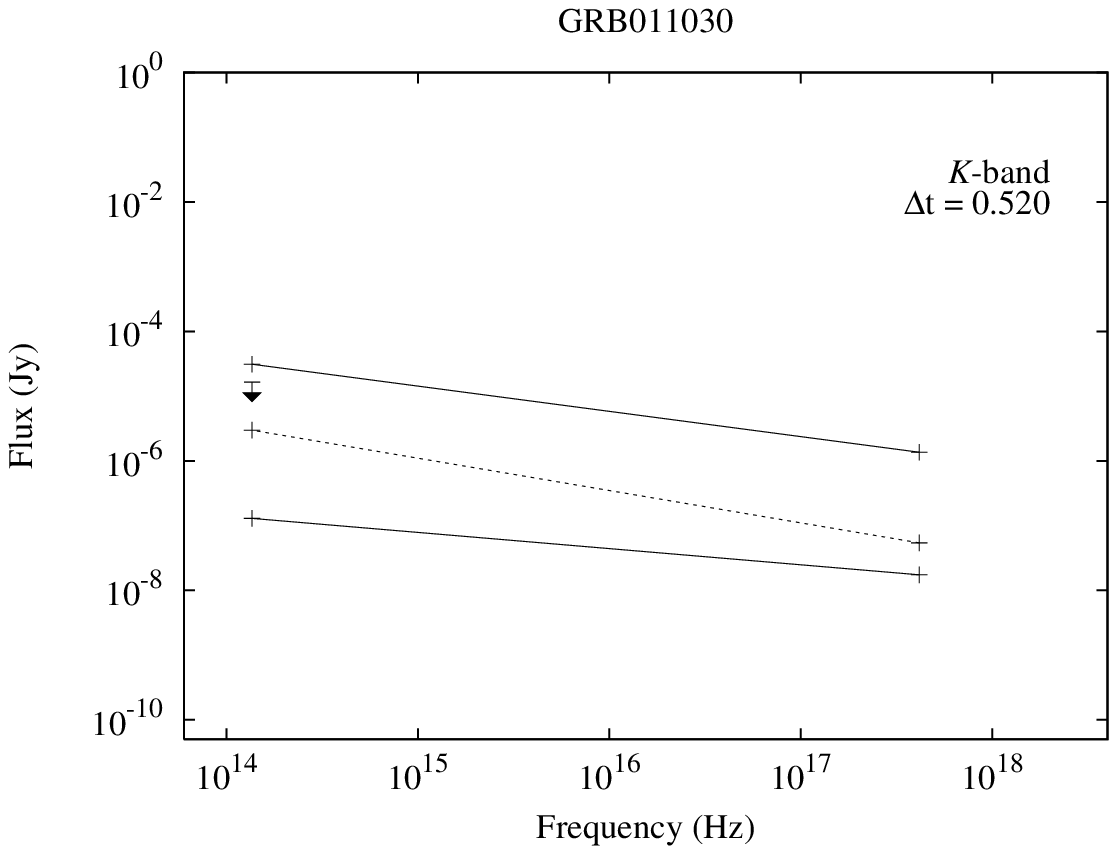}\\[5mm]
    \end{center}
  \end{minipage}
  \begin{minipage}{.45\columnwidth}
    \begin{center}
      (t)
      \includegraphics[width=\columnwidth]{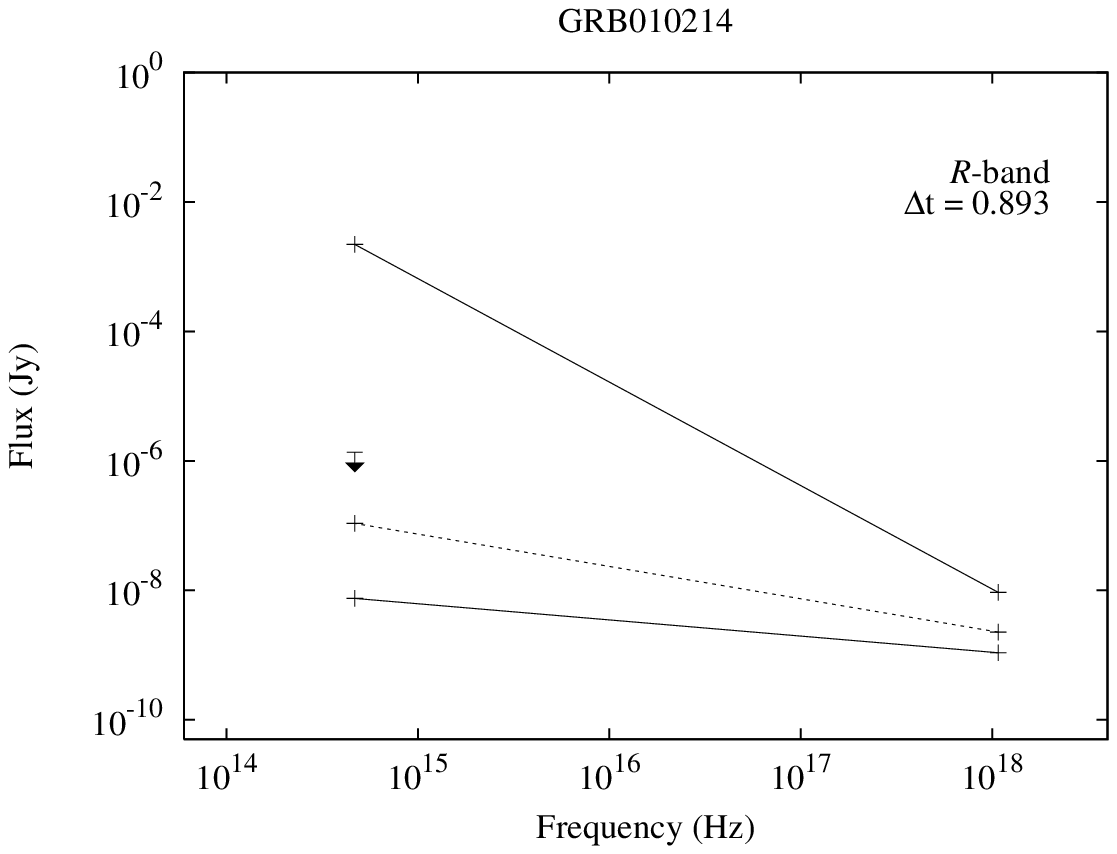}\\[5mm]
      (v)
      \includegraphics[width=\columnwidth]{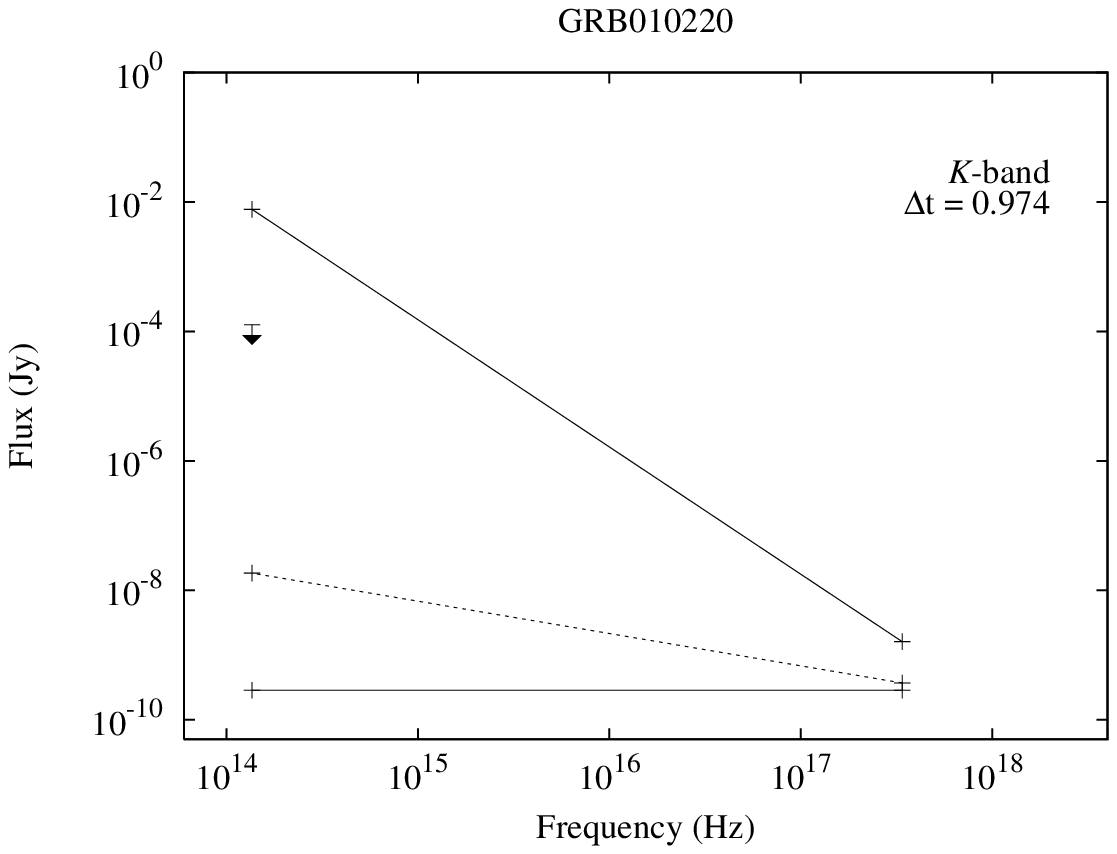}\\[5mm]
      (x)
      \includegraphics[width=\columnwidth]{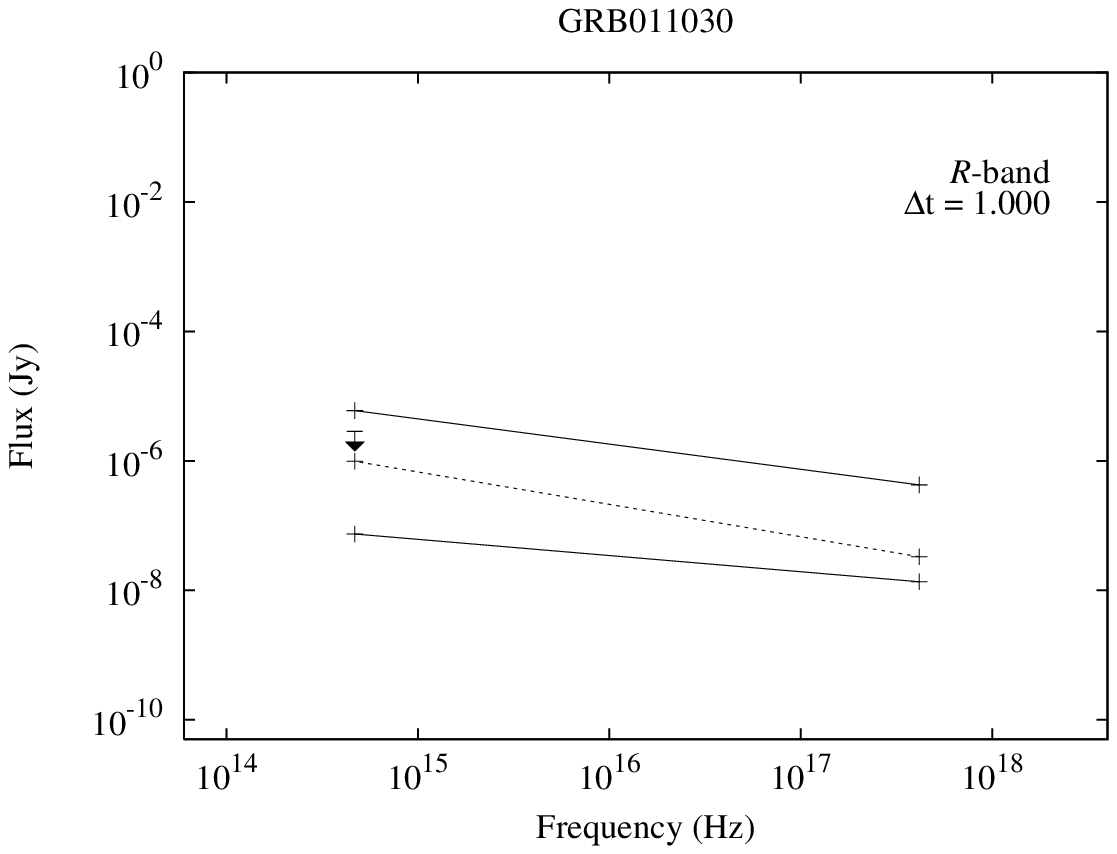}\\[5mm]
    \end{center}
  \end{minipage}

  \figurenum{1}
  \figcaption{(continued)}
 
\end{figure}
\begin{figure}

  \begin{minipage}[t!]{.45\columnwidth}
    \begin{center}
      (y)
      \includegraphics[width=\columnwidth]{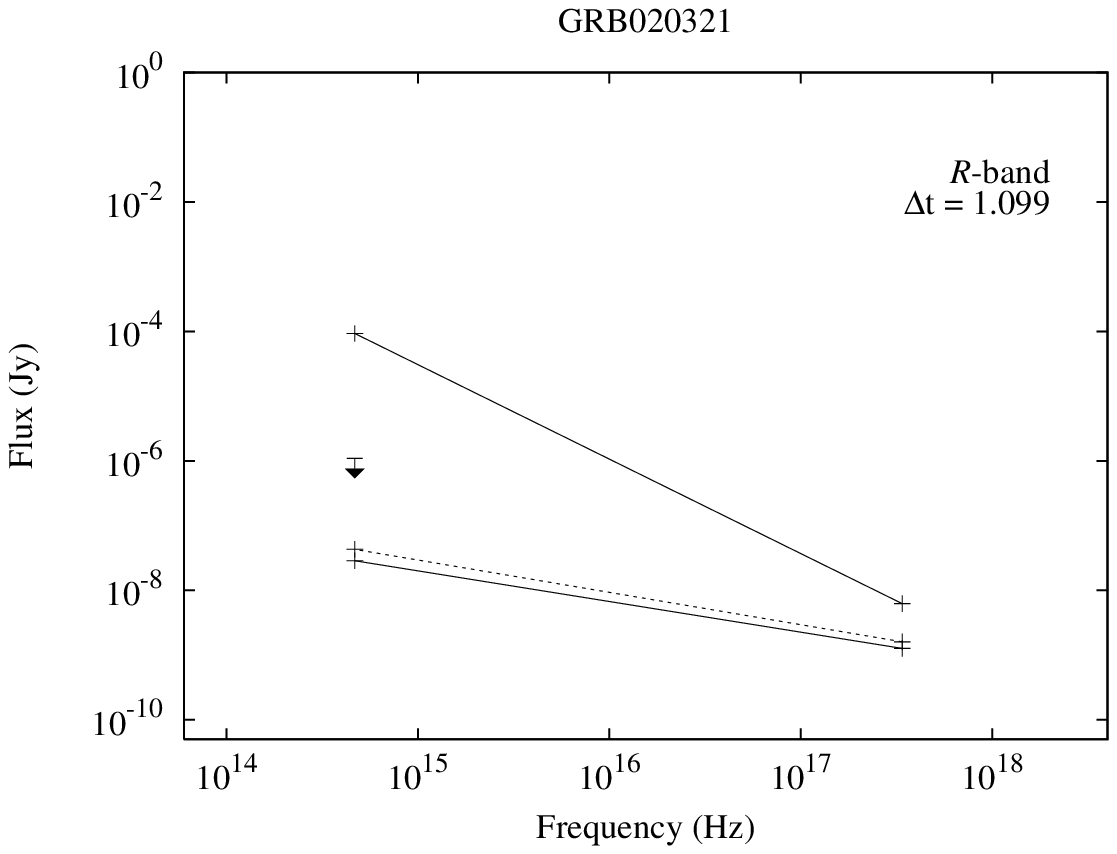}\\[5mm]
      (aa)
      \includegraphics[width=\columnwidth]{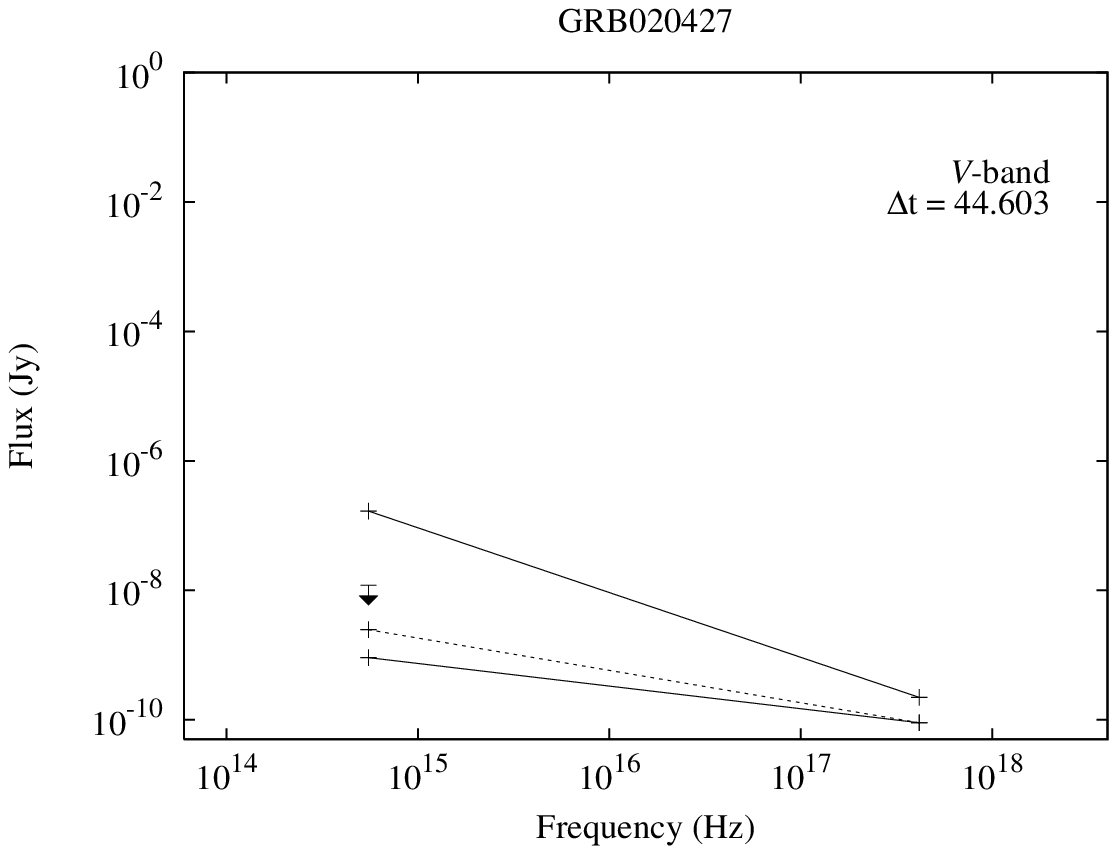}
    \end{center}
  \end{minipage}
  \begin{minipage}[t!]{.45\columnwidth}
    \begin{center}
      (z)
      \includegraphics[width=\columnwidth]{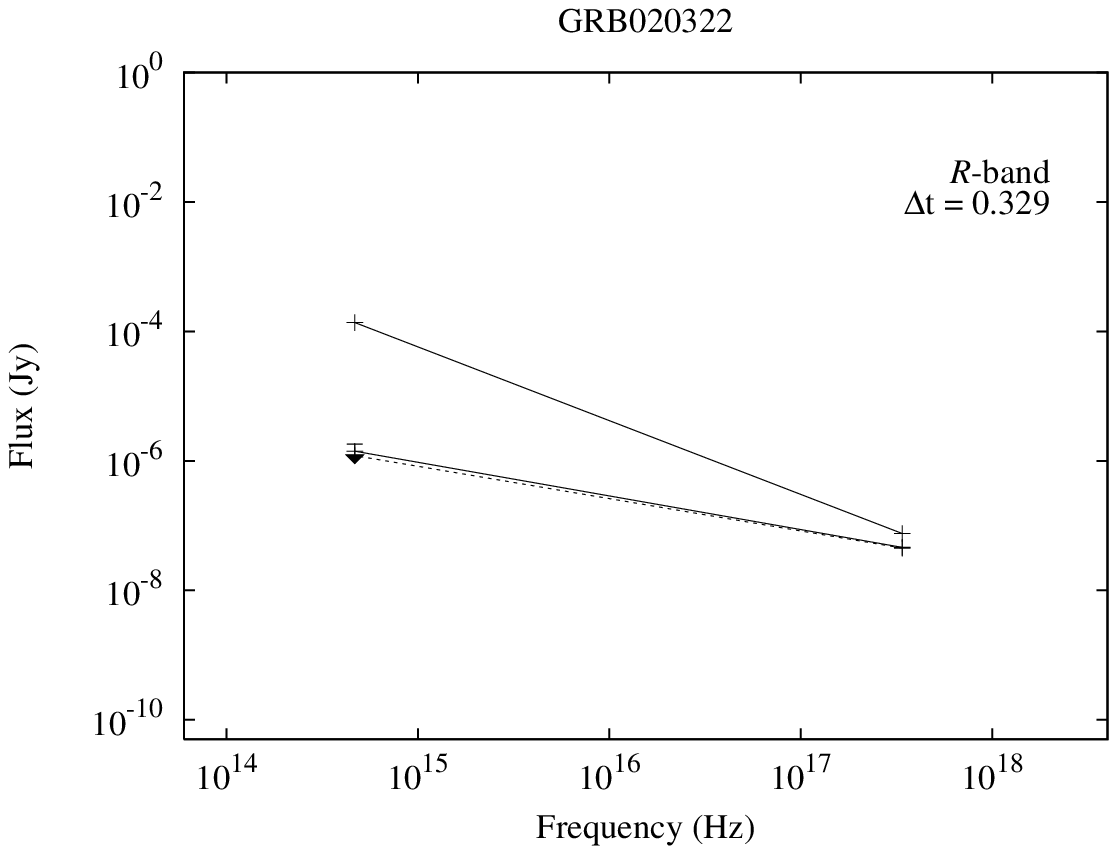}\\[12mm]

      \includegraphics[width=\columnwidth]{f1ab}\\[5mm]
    \end{center}
  \end{minipage}

  \figurenum{1}
  \figcaption{(continued)}
 
\end{figure}

\addtocounter{figure}{1}

\begin{figure}

  \includegraphics[height=0.7\columnwidth]{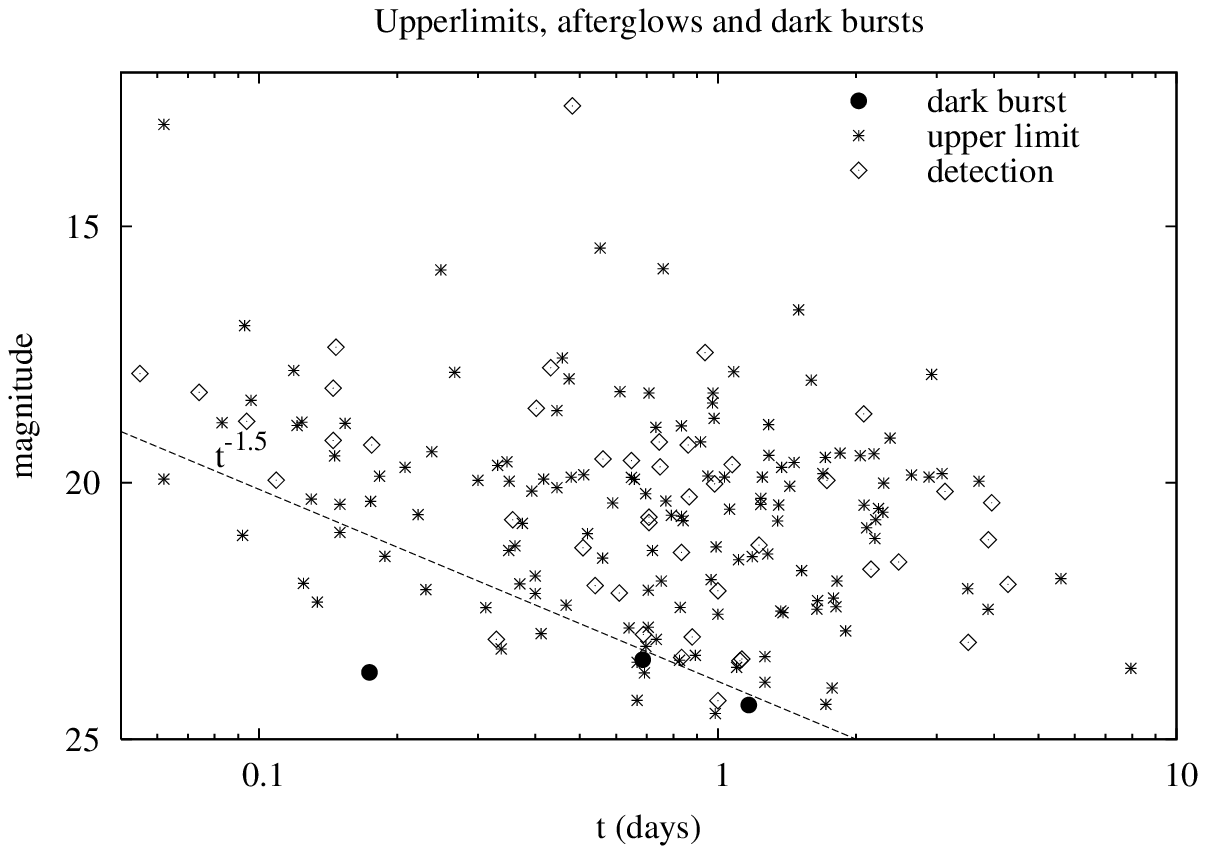}

  \figcaption{\label{figure:darkbursts-magtime}An R-magnitude--versus--time diagram, containing all the $R$-band upper limits (asterisks) from Table \ref{table:darkbursts-upperlimits}, as well as all known detected afterglows (diamonds).
For most afterglows, we have taken the first available $R$-band data point. The dark bursts listed in table \ref{table:darkbursts-extinction} are indicated by the three filled circles. The power law with $\alpha = -1.5$ indicates the region that separates the dark bursts from almost all detected afterglows.}

\end{figure}

\begin{figure}
  \begin{center}
    \includegraphics[width=.8\columnwidth, angle=0]{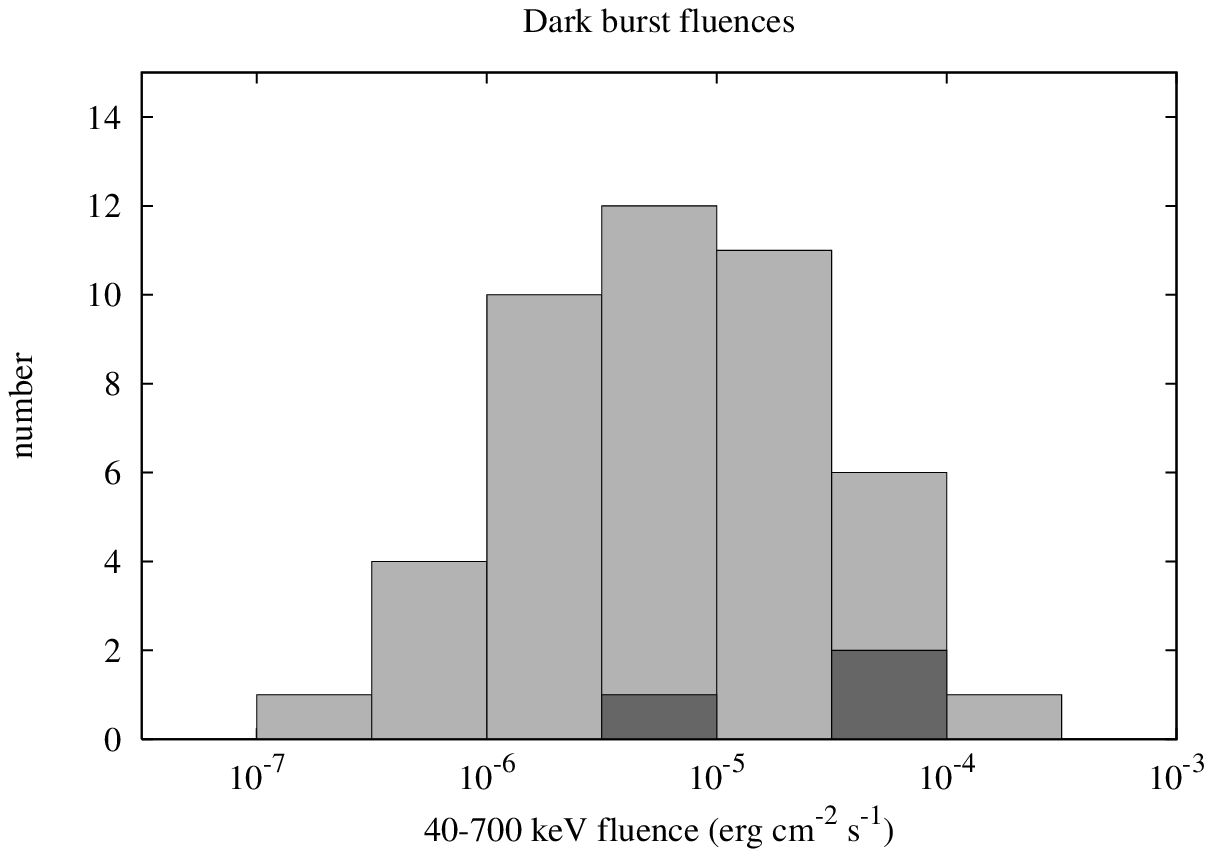}\\

    \includegraphics[width=.8\columnwidth, angle=0]{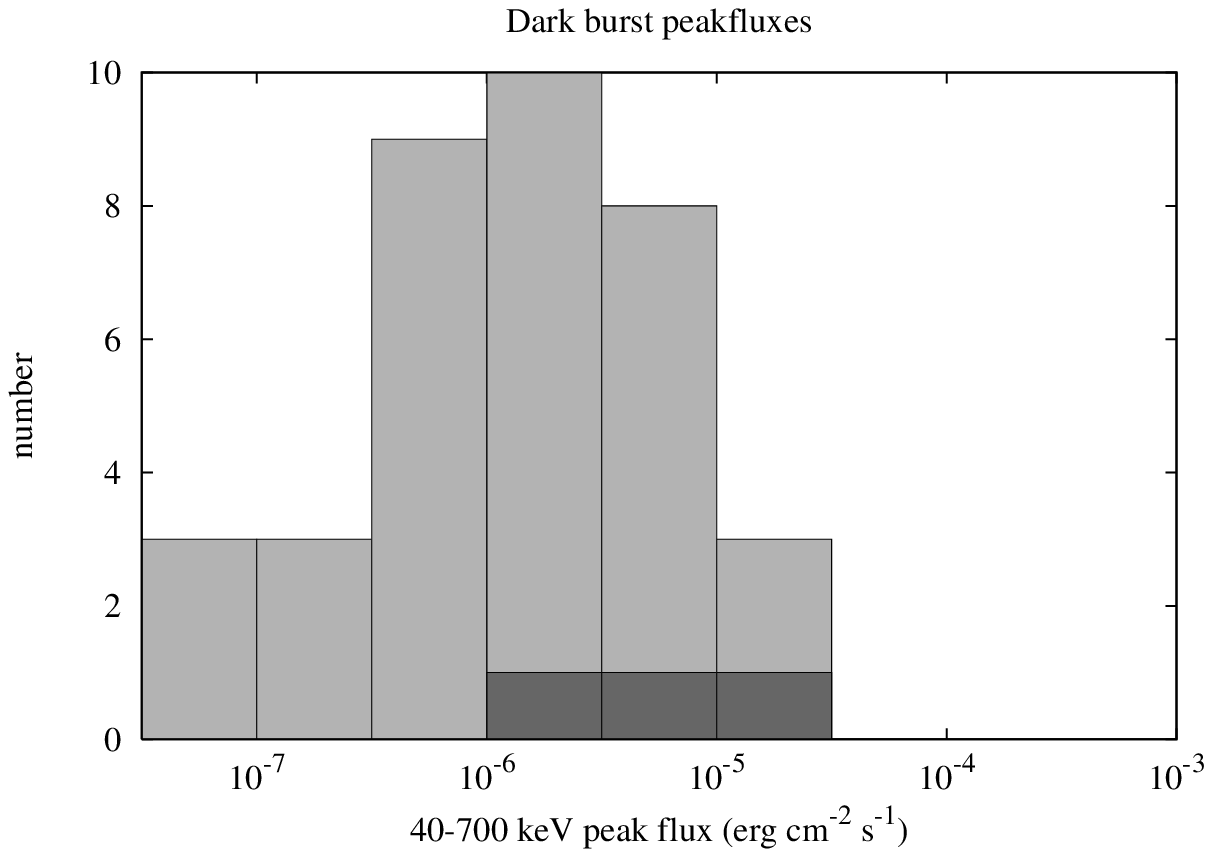}
  \end{center}
  \figcaption{\label{figure:darkbursts-darkpeakflux/fluences} \emph{top} Distribution of the dark bursts fluences (40--700 keV), compared to those of a sample of non-dark BATSE and BeppoSAX bursts. The dark burst histogram is hatched. \emph{bottom} Same as above, but for the 40--700 keV peak fluxes.}

\end{figure}

\begin{deluxetable}{lcccccc}

\tablecaption{Overview of X-ray afterglows of bursts with optical non-detections. We have also indicated the minimum and maximum values of the electron index $p$, which correspond to the minimum and maximum extrapolations in Figure \ref{figure:darkbursts-xray-extrapolations} respectively.\label{table:darkbursts-xrays}}

\tablehead{
  \colhead{burst} &
  \colhead{$\Delta t$} &
  \colhead{flux} &
  \colhead{$\delta$} &
  \colhead{$\beta$} &
  \colhead{$p_{\mathrm{min}}$ -- $p_{\mathrm{max}}$} &
  \colhead{references}\\
  \colhead{} &
  \colhead{(days)} &
  \colhead{($10^{-8}$ Jy)} &
  \colhead{} &
  \colhead{} &
  \colhead{} \\
}

\startdata

  GRB{\,}970402  & 0.620 & $0.69^{+0.13}_{-0.19}$  & $1.57^{+0.03}_{-0.03}$ & $0.70^{+0.60}_{-0.60}$ & $1.54$ -- $3.13$ & (1, 2) \\
  GRB{\,}970828  & 1.571 & $0.99^{+0.23}_{-0.22}$  & $1.44^{+0.07}_{-0.07}$ & $1.60^{+0.20}_{-0.30}$ & $2.60$ -- $2.68$ & (3) \\
  GRB{\,}971227  & 0.723 & $0.75^{+0.15}_{-0.16}$  & $1.12^{+0.08}_{-0.05}$ & $1.10^{+0.22}_{-0.22}$ & $2.09$ -- $2.27$ & (4) \\
  GRB{\,}980613 & 0.458 & $1.20^{+0.49}_{-0.50}$ & $1.19^{+0.17}_{-0.17}$ & $1.10^{+0.22}_{-0.22}$ &  $2.03$ -- $2.81$ & (5) \\
  GRB{\,}981226  & 0.928 & $1.30^{+1.10}_{-1.10}$  & $1.31^{+0.44}_{-0.39}$ & $0.92^{+0.47}_{-0.47}$ & $0.92$ -- $2.78$ & (6) \\
  GRB{\,}990704  & 0.660 & $1.40^{+0.07}_{-0.13}$  & $1.31^{+0.26}_{-0.26}$ & $0.69^{+0.60}_{-0.34}$ & $1.05$ -- $2.58$ & (7) \\
  GRB{\,}990806  & 0.695 & $0.73^{+0.20}_{-0.25}$  & $1.15^{+0.03}_{-0.03}$ & $1.16^{+0.50}_{-0.61}$ & $1.12$ -- $2.57$ & (8) \\
  GRB{\,}991014  & 0.852 & $0.91^{+0.10}_{-0.12}$  & $1.00^{+0.20}_{-0.20}$ & $0.53^{+0.25}_{-0.25}$ & $0.80$ -- $2.56$ & (9) \\
  GRB{\,}000210  & 0.458 & $2.30^{+0.45}_{-0.46}$  & $1.38^{+0.03}_{-0.03}$ & $0.95^{+0.15}_{-0.15}$ & $2.80$ -- $2.88$ & (10) \\
  GRB{\,}000214  & 0.923 & $0.17^{+0.05}_{-0.05}$  & $1.41^{+0.03}_{-0.03}$ & $1.20^{+0.30}_{-0.30}$ & $2.51$ -- $2.92$ & (11) \\
  GRB{\,}000528 & 0.729 & $0.25^{+0.06}_{-0.07}$ & $1.33^{+0.13}_{-0.13}$ & $0.95^{+0.19}_{-0.19}$ &  $2.27$ -- $2.95$ & (12) \\
  GRB{\,}000615  & 0.880 & $0.47^{+0.10}_{-0.11}$  & $1.10^{+0.22}_{-0.22}$ & $1.10^{+0.22}_{-0.22}$ & $1.84$ -- $2.43$ & (13) \\
  GRB{\,}001025A & 2.078 & $0.28^{+0.08}_{-0.13}$  & $3.00^{+1.90}_{-1.90}$ & $1.50^{+0.47}_{-0.12}$ & $2.76$ -- $4.90$ & (14) \\
  GRB{\,}001109  & 0.458 & $11.00^{+3.30}_{-2.70}$ & $2.40^{+0.48}_{-0.48}$ & $1.26^{+0.12}_{-0.49}$ & $1.92$ -- $2.76$ & (15) \\
  GRB{\,}010214  & 0.458 & $1.50^{+0.27}_{-0.64}$  & $2.10^{+1.00}_{-0.60}$ & $0.30^{+0.80}_{-0.60}$ & $1.50$ -- $3.20$ & (16, 15) \\
  GRB{\,}010220  & 0.829 & $0.12^{+0.00}_{-0.07}$  & $1.20^{+1.00}_{-1.00}$ & $1.10^{+1.00}_{-0.60}$ & $1.00$ -- $3.93$ & (14) \\
  GRB{\,}011030 & 10.726 & $0.59^{+0.03}_{-0.03}$ & $2.10^{+0.60}_{-0.60}$ & $0.31^{+0.08}_{-0.08}$ & $1.50$ -- $1.78$ & (17) \\
  GRB{\,}020321  & 0.742 & $0.75^{+0.37}_{-0.40}$  & $1.20^{+0.24}_{-0.24}$ & $1.10^{+0.22}_{-0.22}$ & $1.95$ -- $2.92$ & (18) \\
  GRB{\,}020322\tablenotemark{a}  & 0.751 & $2.60^{+0.16}_{-0.18}$  & $1.26^{+0.23}_{-0.23}$ & $1.06^{+0.08}_{-0.08}$ & $2.04$ -- $2.28$ & (19) \\
  GRB{\,}020427 & 9.164 & $0.27^{+0.03}_{-0.06}$ & $2.30^{+0.60}_{-0.60}$ & $0.30^{+0.20}_{-0.20}$ & $1.70$ -- $2.00$ & (17) \\
  GRB{\,}040223 & 0.393 & $1.1^{+0.3}_{-0.3}$ & $1.0^{+0.2}_{-0.2}$ & $ 1.8^{+0.2}_{-0.2}$ & ---\tablenotemark{b} & (20) \\

\enddata  
\tablenotetext{a}{Optically detected burst. See Sect. \ref{section:darkbursts-results} for details.}
\tablenotetext{b}{The values for $\delta$ and $\beta$ are incompatible with each other within the variants of the fireball model used in this paper, yielding no possible range for $p$}
\tablerefs{
(1) \citet{nicastro1998:aa338};
(2) \citet{piro1997:iauc6617};
(3) \citet{yoshida2001:apj557};
(4) \citet{antonelli1999:aas138};
(5) \citet{costa1998:aas193};
(6) \citet{frontera2000:apj540};
(7) \citet{feroci2001:aa378};
(8) \citet{montanari2001:proc};
(9) \citet{intzand2000:apj545};
(10) \citet{piro2002:apj577};
(11) \citet{antonelli2000:apj545};
(12) \citet{frontera2003:proc};
(13) \citet{nicastro2001:proc};
(14) \citet{watson2002:aa393};
(15) \citet{depasquale2003:apj592};
(16) \citet{guidorzi2003:aa401};
(17) Levan et al, submitted;
(18) \citet{intzand2003:proc};
(19) \citet{watson2002:aa395};
(20) \citet{tiengo2004:gcn2548}
}

\end{deluxetable}

\begin{deluxetable}{lccccl}
\tablecaption{Overview of optical/near-IR upper limits with X-ray afterglow data available\label{table:darkbursts-upperlimits}}
\tablehead{
    \colhead{burst} &
	\colhead{filter} &
	\colhead{$\Delta t$} &
	\colhead{magnitude} &
	\colhead{E(B-V)} &
	\colhead{reference}\\
	\colhead{} &
	\colhead{} &
	\colhead{(days)} &
	\colhead{(uncorrected)} &
	\colhead{} &
	\colhead{} \\
    }

\startdata
GRB{\,}970402	& R	& 0.770	& 21.00	& 0.24	& \citet{groot1997:iauc6616}  \\
GRB{\,}970828	& R	& 0.174	& 23.80	& 0.04	& \citet{groot1998:apj493}  \\
GRB{\,}971227	& R	& 0.561	& 21.50	& 0.01	& \citet{castro-tirado1997:gcn18}  \\
GRB{\,}981226	& I	& 0.692	& 23.50	& 0.02	& \citet{bloom1998:gcn182}  \\
GRB{\,}981226	& R	& 0.412	& 23.00	& 0.02	& \citet{lindgren1999:gcn190}  \\
GRB{\,}990704	& R	& 0.231	& 22.20	& 0.04	& \citet{rol1999:gcn374}  \\
GRB{\,}990806	& B	& 0.667	& 25.20	& 0.04	& \citet{greiner1999:gcn396}  \\
GRB{\,}990806	& R	& 0.697	& 23.30	& 0.04	& \citet{kemp1999:gcn397}  \\
GRB{\,}991014	& R	& 0.467	& 23.10	& 0.27	& \citet{thorstensen1999:gcn423}  \\
GRB{\,}991014	& K	& 1.240	& 19.00	& 0.27	& \citet{klose2003:apj592}  \\
GRB{\,}000210	& R	& 0.686	& 23.50	& 0.02	& \citet{gorosabel2000:gcn783}  \\
GRB{\,}000214	& R	& 0.337	& 23.24	& 0.06	& this paper  \\
GRB{\,}000214	& K	& 1.350	& 18.10	& 0.06	& \citet{rhoads2000:gcn564}  \\
GRB{\,}000528	& R	& 0.734	& 23.30	& 0.09	& \citet{palazzi2000:gcn691}  \\
GRB{\,}000615	& R	& 0.188	& 21.50	& 0.02	& \citet{stanek2000:gcn709}  \\
GRB{\,}000615	& H	& 0.705	& 20.50	& 0.02	& \citet{pian2000:gcn727}  \\
GRB{\,}001025A	& R	& 1.168	& 24.50	& 0.06	& \citet{fynbo2000:gcn867}  \\
GRB{\,}001109	& R	& 0.375	& 20.90	& 0.04	& \citet{greiner2000:gcn887}  \\
GRB{\,}001109	& K	& 0.400	& 19.90	& 0.04	& \citet{castro-ceron2004:aa424}  \\
GRB{\,}010214	& R	& 0.893	& 23.50	& 0.05	& \citet{rol2001:gcn1059}  \\
GRB{\,}010220	& R	& 0.361	& 23.50	& 0.85	& \citet{berger2001:gcn958}  \\
GRB{\,}010220	& K	& 0.974	& 17.10	& 0.85	& \citet{licausi2001:gcn979}  \\
GRB{\,}011030	& R	& 0.300	& 21.00	& 0.39	& \citet{mohan2001:gcn1120}  \\
GRB{\,}011030	& K	& 0.520	& 19.00	& 0.39	& \citet{klose2001:gcn1142}  \\
GRB{\,}011030	& R	& 1.000	& 23.61	& 0.39	& \citet{rhoads2001:gcn1140}  \\
GRB{\,}011030	& V	& 42.905 & 27.6	& 0.39	& \citet{levan2004:apj}  \\
GRB{\,}020321	& R	& 1.099	& 23.60	& 0.10	& \citet{salamanca2002:gcn1385}  \\
GRB{\,}020322	& R	& 0.329	& 23.26	& 0.08	& \citet{bloom2002:gcn1294}\tablenotemark{a}	\\
GRB{\,}020427	& V	& 44.603 & 28.7	& 0.03	& \citet{levan2004:apj}  \\

\enddata
\tablenotetext{a}{Optically detected burst. See Sect. \ref{section:darkbursts-results} for details.}

\end{deluxetable}

\begin{deluxetable}{lcccc}

\tablecaption{\label{table:darkbursts-powerlaw-indices} The temporal ($\delta$) and spectral ($\beta$) indices as a function of the electron power-law distribution index $p$ in the slow cooling case, for eight different scenarios. See e.g. \citet{sari1998:apj497} and \citet{chevalier1999:apj520}.}

\tablehead{
  \colhead{} &
  \colhead{$\delta (p)$} &
  \colhead{$\beta (p)$} &
  \colhead{$\delta (p)$} &
  \colhead{$\beta (p)$}
}

  \startdata

  &
  \multicolumn{2}{c}{$\nu > \nu_c$} &
  \multicolumn{2}{c}{$\nu < \nu_c$} \\

  \sidehead{Pre jet-break (isotropic)} 
  \hline
  ISM ($\rho = \mathrm{constant}$) & 
  $(3p-2)/4$ & $p/2$     & 
  $3(p-1)/4$ & $(p-1)/2$ \\
  
  Stellar wind ($\rho \propto R^{-2}$) &
  $(3p-2)/4$ & $p/2$     &
  $(3p-1)/4$ & $(p-1)/2$ \\

  \sidehead{Post jet-break (collimated)} 
  \hline
  ISM ($\rho = \mathrm{constant}$) & 
  $p$ & $p/2$     & 
  $p$ & $(p-1)/2$ \\
  
  Stellar wind ($\rho \propto R^{-2}$) &
  $p$ & $p/2$     & 
  $p$ & $(p-1)/2$ \\
  
  \enddata

\tablecomments{ We have chosen $\delta$ and $\beta$ in such a way that they are positive in optical and X-rays for positive $p$, i.e. the afterglow flux follows $F (\nu, t) \propto \nu^{-\beta} t^{-\delta}$. }

\end{deluxetable}

\begin{deluxetable}{lcccc}

\tabletypesize{\scriptsize}

\tablecaption{Burst classification and amount of (host galaxy) extinction inferred (in the observer frame) \label{table:darkbursts-extinction}}

\tablehead{
  \colhead{burst} &
  \colhead{filters} &
  \colhead{classification} &
  \colhead{extinction (mag)} \\
}

  \startdata

 GRB970402 & $R$ & grey \\[2mm]
 GRB970828 & $R$ & dark &  $ \ge 4.304$ \\[2mm]
 GRB971227 & $R$ & grey \\[2mm]
 GRB981226 & $R$ & grey \\[2mm]
 GRB990704 & $R$ & grey \\[2mm]
 GRB990806 & $B$ & grey \\[2mm]
 GRB991014 & $K$, $R$ & grey \\[2mm]
 GRB000210 & $R$ & dark &  $ \ge 1.708$ \\[2mm]
 GRB000214 & $K$, $R$ & grey \\[2mm]
 GRB000528 & $R$ & grey \\[2mm]
 GRB000615 & $H$, $R$ & grey \\[2mm]
 GRB001025A & $R$ & dark &  $ \ge 0.690$ \\[2mm]
 GRB001109 & $K$, $R$ & grey \\[2mm]
 GRB010214 & $R$ & grey \\[2mm]
 GRB010220 & $K$, $R$ & grey \\[2mm]
 GRB020321 & $K$, $R$ & grey \\[2mm]
  
  \enddata

\end{deluxetable}

\end{document}